\documentclass[journal=jctcce,manuscript=article,layout=twocolumn]{achemso}

\usepackage[version=3]{mhchem} 
\usepackage{hyperref}
\usepackage{pdfpages} 

\usepackage{xcolor}

\let\oldmaketitle\maketitle
\let\maketitle\relax

\author{Michele Invernizzi}
\affiliation{Freie Universit\"at Berlin, 14195 Berlin, Germany}
\email{michele.invernizzi@fu-berlin.de}
\author{Michele Parrinello}
\affiliation{Italian Institute of Technology, 16163 Genova, Italy}

\title{Exploration vs Convergence Speed\\in Adaptive-bias Enhanced Sampling}

\abbreviations{OPES, MetaD, CVs, FES}

\begin{document}

\newlength{\myfigwidth}
\setlength{\myfigwidth}{\columnwidth}






\twocolumn[
\begin{@twocolumnfalse}
\oldmaketitle
\begin{abstract}
In adaptive-bias enhanced sampling methods, a bias potential is added to the system to drive transitions between metastable states.
The bias potential is a function of a few collective variables and is gradually modified according to the underlying free energy surface.
We show that when the collective variables are suboptimal, there is an exploration-convergence tradeoff, and one must choose between a quickly converging bias that will lead to fewer transitions, or a slower to converge bias that can explore the phase space more efficiently but might require a much longer time to produce an accurate free energy estimate.
The recently proposed On-the-fly Probability Enhanced Sampling (OPES) method focuses on fast convergence, but there are cases where fast exploration is preferred instead.
For this reason, we introduce a new variant of the OPES method that focuses on quickly escaping metastable states, at the expense of convergence speed.
We illustrate the benefits of this approach on prototypical systems and show that it outperforms the popular metadynamics method.
\end{abstract}
\end{@twocolumnfalse}
]
\section{Introduction}
Molecular dynamics has become a valuable tool in the study of a variety of phenomena in physics, chemistry, biology, and materials science.
One of the long-standing challenges in this important field is the sampling of rare events, such as chemical reactions or conformational changes in biomolecules.
To simulate effectively such systems, many enhanced sampling methods have been developed.
An important class of such methods is based on an adaptive-bias approach and includes adaptive umbrella sampling\cite{Mezei1987}, metadynamics (MetaD) \cite{Laio2002,Barducci2008}, and the recently developed on-the-fly probability enhanced sampling (OPES)\cite{Invernizzi2020rethinking,Invernizzi2020unified,Invernizzi2021}.
Adaptive-bias methods operate by adding to the system's energy $U(\mathbf{R})$ an external bias potential $V=V(\mathbf{s})$, that is a function of a set of collective variables (CVs), $\mathbf{s}$.
The CVs, $s=s(\mathbf{R})$, depend on the atomic coordinates $\mathbf{R}$ and are meant to describe the slow modes associated with the rare event under study.
They also define a free energy surface (FES), $F(\mathbf{s})=-\frac{1}{\beta}\log P(\mathbf{s})$, where $\beta=(k_{B}T)^{-1}$ is the inverse Boltzmann factor and $P(\mathbf{s})$ the marginal $\mathbf{s}$ distribution, $P(\mathbf{s}) \propto \int e^{-\beta U(\mathbf{R})} \delta[\mathbf{s}-\mathbf{s}(\mathbf{R})] \,d \mathbf{R}$.
The bias is periodically updated until it converges to a chosen form.
A popular choice is to have it exactly offset the underlying FES, $V(\mathbf{s})=-F(\mathbf{s})$, so that the resulting $\mathbf{s}$ distribution is uniform.

The main limitation of adaptive-bias methods is that finding good collective variables is sometimes difficult and a bad choice of CVs might not promote the desired transitions in an affordable computer time.
In practical applications one generally has to live with suboptimal CVs\cite{Invernizzi2019} that still can drive transitions, but do not include some of the slow modes.
In this case, applying a static bias cannot speed up the slow modes that are not accounted for, and thus transitions remain quite infrequent.
It is sometimes possible to achieve a faster transition rate by using a rapidly changing bias, which can push the system out of a metastable state through a high free energy pathway, different from the energetically favoured one.
However, unless one wishes to deal explicitly  with out-of-equilibrium statistics\cite{Jarzynski1997,Donati2018,Bal2021}, it is not possible to obtain reliable information about the system while the bias changes in a non-adiabatic fashion.
To estimate the FES and other observables one must let the adaptive-bias method approach convergence, and as the bias becomes quasi-static, transitions inevitably become less frequent.

We refer to this situation as an exploration-convergence tradeoff that every adaptive-bias enhanced sampling method has to deal with, when suboptimal CVs are used.
Some methods, like OPES, focus more on quickly converging to a quasi-static bias potential and thus obtaining an efficiently reweighted FES, while others, like metadynamics, focus more on escaping metastable states and exploring the phase space.
We will demonstrate this qualitative difference on some prototypical systems.
For simplicity, in the paper we only consider the well-tempered variant of metadynamics\cite{Barducci2008}, but in the Supporting Information (SI) we provide examples  that use the original non-tempered MetaD\cite{Laio2002} and other popular variants, such as parallel-bias MetaD\cite{Pfaendtner2015}.

We propose here a variant of  OPES, named OPES-explore, that focuses on rapid exploration, rather than on fast convergence.
It shares many features with the original OPES, and is designed to be an easy-to-use tool requiring few input parameters.
To this end, we also introduce an adaptive bandwidth algorithm that can be used in both OPES variants, and further reduces the number of input parameters that need to be specified.
The detailed description of the adaptive bandwidth algorithm is left to the SI.
All OPES simulations presented make use of this algorithm.

\section{The OPES method}\label{S:opes}
The enhanced sampling method OPES works by adding an adaptive-bias potential to the energy of the system, so as to modify the Boltzmann probability distribution into a desired target one.
Most adaptive-bias methods aim at sampling uniformly the CV space, but it has been shown that choosing a different target distribution could be advantageous\cite{Valsson2014,White2015}.
There are two different classes of target distributions that can be sampled with OPES; metadynamics-like and replica-exchange-like.
We will consider here only the former type, introduced in Ref.~\citenum{Invernizzi2020rethinking}, but the interested reader can find in Ref.~\citenum{Invernizzi2020unified} information about OPES for replica-exchange-like sampling.

To define a metadynamics-like target distribution, one has to choose a set of collective variables, $s=s(\mathbf{R})$.
As stated in the introduction, the unbiased marginal probability along such CVs is $P(\mathbf{s}) \propto \int e^{-\beta U(\mathbf{R})} \delta[\mathbf{s}-\mathbf{s}(\mathbf{R})] \,d \mathbf{R}$, where $U(\mathbf{R})$ is the potential energy.
The target distribution is then defined by requiring a specific marginal probability distribution over the CVs, $p^{\text{tg}}(\mathbf{s})$.
Consequently, the desired bias potential is written as:
\begin{equation}\label{E:bias}
    V(\mathbf{s})=-\frac{1}{\beta} \log \frac{p^{\text{tg}}(\mathbf{s})}{P(\mathbf{s})}\, ,
\end{equation}
so that $\int e^{-\beta [U(\mathbf{R})+ V(\mathbf{s})]} \delta[\mathbf{s}-\mathbf{s}(\mathbf{R})] \,d \mathbf{R} \propto p^{\text{tg}}(\mathbf{s})$.
A typical choice for $p^{\text{tg}}(\mathbf{s})$ is the well-tempered distribution\cite{Barducci2008}:
\begin{equation}\label{E:well-tempered}
    p^{\text{WT}}(\mathbf{s}) \propto [P(\mathbf{s})]^{1/\gamma}\, ,
\end{equation}
where the bias factor $\gamma>1$ controls how much the original distribution is smoothed out.
In the limit of $\gamma=\infty$ one targets a uniform distribution.

The core idea of OPES is to update self-consistently the estimate of the probability distributions and of the bias potential, in an on-the-fly fashion similar to self-healing umbrella sampling \cite{Marsili2006}.
The estimate of the unbiased probability is obtained via a weighted kernel density estimation, so that at step $n$ one has:
\begin{equation}\label{E:iter_prob}
    P_n(\mathbf{s})=\frac{\sum_k^n w_k G(\mathbf{s},\mathbf{s}_k)}{\sum_k^n w_k}\, ,
\end{equation}
where the weights $w_k$ are given by $w_k=e^{\beta V_{k-1}(\mathbf{s}_k)}$, and the Gaussian kernels $G(\mathbf{s},\mathbf{s}')=h\exp \left[-\frac{1}{2} (\mathbf{s}-\mathbf{s}')^T\boldsymbol{\Sigma}^{-1} (\mathbf{s}-\mathbf{s}') \right]$ have a diagonal covariance matrix $\Sigma_{ij}=\sigma^2_i\delta_{ij}$ and fixed height $h=\prod_i \left(\sigma_i\sqrt{2\pi}\right)^{-1}$.
The number of kernels to represent $P_n(\mathbf{s})$ would grow linearly with simulation time, but this is avoided thanks to an on-the-fly kernel compression algorithm\cite{Sodkomkham2016}, as described in detail in the supporting information of Ref.~\citenum{Invernizzi2020rethinking}.
The compression algorithm also allows for the bandwidth of the kernels to shrink over time, as the effective sample size $N_{\text{eff}}^{(n)}=\left(\sum_k^n w_k\right)^2/\sum_k^n w_k^2$ grows.
The idea is to start with a coarse estimate of $P(\mathbf{s})$ and then refine it as more data are available.
The kernel bandwidth of the $i$-th CV at step $n$ is:
\begin{equation}\label{E:bandwidth}
    \sigma_i^{(n)}=\sigma_i^{(0)} [N_{\text{eff}}^{(n)}(d+2)/4]^{-1/(d+4)}\, ,
\end{equation}
where $d$ is the total number of CVs.

The instantaneous bias is based on the probability estimate $P_n(\mathbf{s})$, following Eq.~(\ref{E:bias}) and using the approximation $p^{\text{WT}}(\mathbf{s}) \propto [P_n(\mathbf{s})]^{1/\gamma}$, one has:
\begin{equation}\label{E:iter_bias}
    V_n(\mathbf{s})=(1-1/\gamma)\frac{1}{\beta} \log \left( \frac{P_n(\mathbf{s})}{Z_n}+\epsilon \right)\, ,
\end{equation}
where $\epsilon$ is a regularization term that limits the maximum possible absolute value of the bias potential, and $Z_n$ can be understood as a normalization of $P_n(\mathbf{s})$ over the CV space thus far explored, $\Omega_n$:
\begin{equation}\label{E:iter_zed}
    Z_n=\frac{1}{|\Omega_n|}\int_{\Omega_n} P_n(\mathbf{s})\, d \mathbf{s}\, .
\end{equation}
This integral is calculated approximately as a sum of $P_n$ over the compressed kernels, as explained in the supplementary information of Ref.~\citenum{Invernizzi2020rethinking}.
The intuitive idea is that new kernels are added to the compressed representation only when a new region of CV space is sampled (otherwise they are merged with existing ones), thus the explored CV-space volume $|\Omega_n|$, is approximately proportional to the total number of compressed kernels.

The introduction of the $Z_n$ term is one of the key innovations of OPES. 
In similar methods, once a new metastable state is found one often sees a dramatic increase of the exit time, compared to the first one\cite{Fort2017} (see SI, Fig.~S5).
This exit time problem is present also when the CVs are optimal, and should not be confused with the exploration-convergence tradeoff that is the primary concern of this paper.
Other convergence-focused methods introduce extra parameters to tackle this problem, for example in transition-tempered metadynamics\cite{Dama2014_ttmetad} prior knowledge of the position of all metastable states is required.
Instead, OPES avoids the exit time problem by taking into account the expansion of the CV space via the $Z_n$ term, which allows the bias to adjust more quickly when a new CV-space region is sampled\cite{Invernizzi2020rethinking}.

 At the start of an OPES simulation only a handful of parameters needs to be chosen, namely the initial kernel bandwidth, the pace at which the bias is updated, and the approximate FES barrier height that needs to be overcome.
 From this last information a prescription is given to automatically set the values $\gamma$ and $\epsilon$.
The number of parameters  can be reduced even further if one uses, as we shall do here, the adaptive bandwidth algorithm discussed in the SI.

\section{An exploratory OPES variant}

We present now a new OPES variant called OPES-explore which, compared to the original OPES formulation, leads to a faster exploration of the phase space at the cost of a slower convergence.
We have recalled that the $Z_n$ term allows OPES to quickly adapt the bias when a metastable state is found in a previously unexplored region of CV space.
However, if the CVs used are suboptimal, it may happen that a new metastable state is found in an already explored $\mathbf{s}$ region\cite{Pietrucci2017,Bussi2020}.
In such a case, the $Z_n$ term remains constant and is therefore ineffective in accelerating the exit time.
Instead, to encourage a rapid exit, one would need a method that allows the bias to significantly change shape again.
Fortunately, it is possible to achieve this exploratory behaviour simply by making a minimal change to the OPES protocol, which gives rise to the OPES-explore variant.

In formulating OPES-explore, we restrict ourselves to the case of using as target the well-tempered distribution, $p^{\text{tg}}(\mathbf{s})=p^{\text{WT}}(\mathbf{s})$, Eq.~(\ref{E:well-tempered}).
In OPES, the bias is expressed as a function of $P_n(\mathbf{s})$, the on-the-fly estimate of the unknown equilibrium distribution $P(\mathbf{s})$.
At the beginning of the simulation this estimate is not reliable, but it improves over time and converges in a self-consistent way.
In OPES-explore instead, one builds the bias starting from the on-the-fly estimate of the distribution that is being sampled in the biased simulation:
\begin{equation}\label{E:iter_prob-explore}
    p^{\text{WT}}_n(\mathbf{s})=\frac{1}{n} \sum_k^n G(\mathbf{s},\mathbf{s}_k)\, ,
\end{equation}
where $\mathbf{s}_k$ is the CVs value sampled at step $k$.
As the simulation converges, $p^{\text{WT}}_n(\mathbf{s})$ approaches the target well-tempered distribution $p^{\text{WT}}(\mathbf{s})$.
Thus, analogously to Sec.~\ref{S:opes}, we use the approximation $P(\mathbf{s})\propto [p^{\text{WT}}_n(\mathbf{s})]^\gamma$ and write the bias according to Eq.~(\ref{E:bias}):
\begin{equation}\label{E:iter_bias-explore}
    V_n(\mathbf{s}) = (\gamma-1)\frac{1}{\beta} \log \left( \frac{p^{\text{WT}}_n(\mathbf{s})}{Z_n}+\epsilon \right)\, ,
\end{equation}
where $\epsilon$ and $Z_n$ have been added for the same reasons as in Eq.~($\ref{E:iter_bias}$).
We notice that the expressions in  Eqs.~(\ref{E:iter_prob}) and (\ref{E:iter_prob-explore}), which define the probability estimates used in the two OPES schemes, converge respectively to $P(\mathbf{s})$ and $p^{\text{WT}}(\mathbf{s})$ only within the self-consistent scheme where the simulation runs with a bias that is updated on-the-fly according to Eqs.~(\ref{E:iter_bias}) and (\ref{E:iter_bias-explore}) respectively.
Both OPES variants are applications of the general Eq.~(\ref{E:bias}), but OPES estimates on-the-fly $P(\mathbf{s})$ and uses it to calculate the bias, while OPES-explore does the same but with $p^{\text{WT}}(\mathbf{s}) \propto [P(\mathbf{s})]^{1/\gamma}$.

The free energy surface as a function of the CVs can be estimated in two distinct ways, either directly from the probability estimate, $F_n(\mathbf{s})=-\gamma\frac{1}{\beta}\log p_n^{\text{WT}}(\mathbf{s})$, or via importance sampling reweighting, e.g.~using a weighted kernel density estimation,
\begin{equation}\label{E:reweighting}
    F_n(\mathbf{s})= -\frac{1}{\beta} \log \sum_k^n e^{\beta V_{k-1}(\mathbf{s}_k)} G(\mathbf{s},\mathbf{s}_k)\, .
\end{equation}
In standard OPES these two estimates are equivalent, while in OPES-explore (similarly to MetaD) they can differ significantly in the first part of the simulation until they eventually converge to the same estimate. 

\begin{figure*}
    \begin{tabular}{cc}
        \includegraphics[width=\myfigwidth]{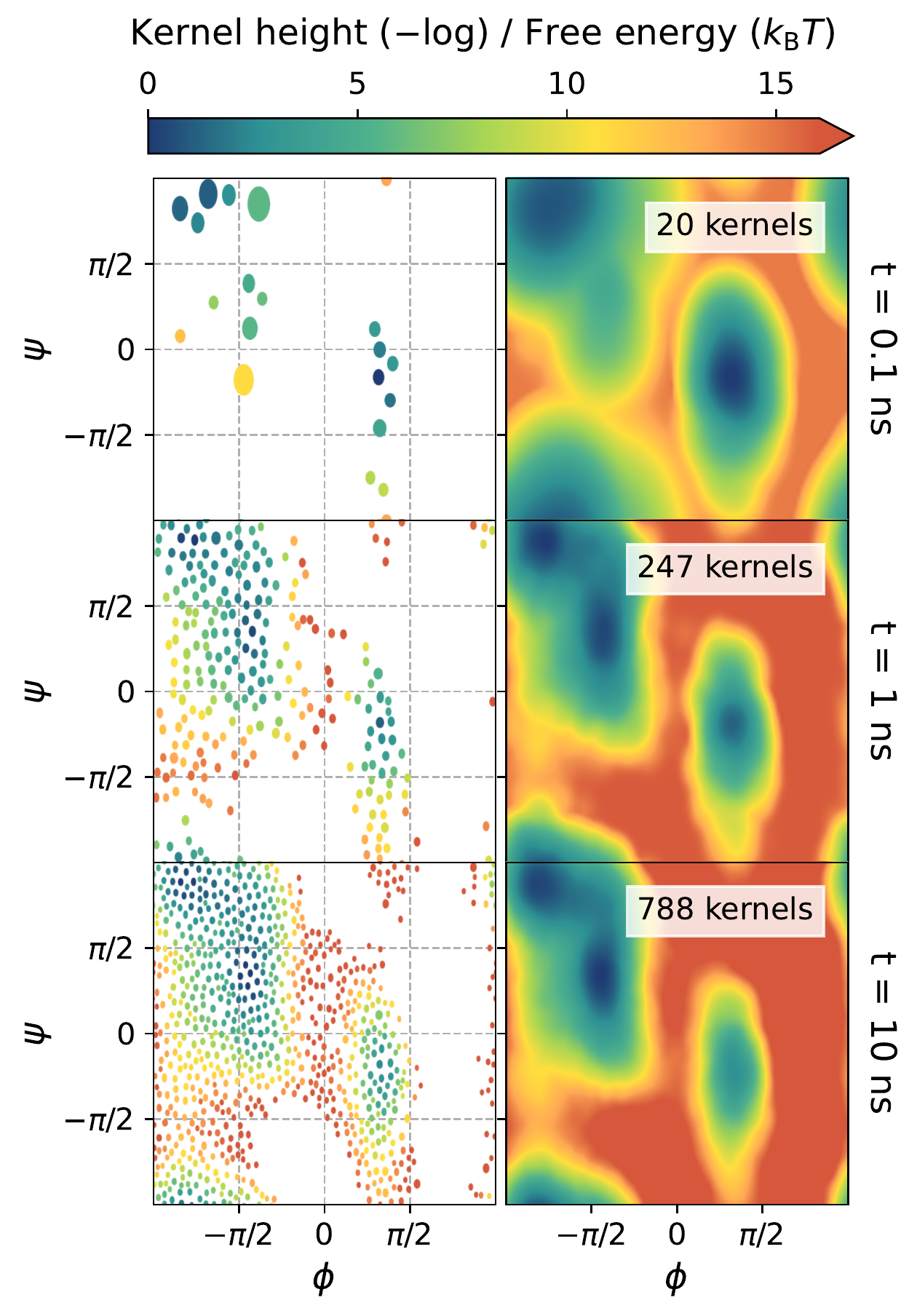}
        & \includegraphics[width=\myfigwidth]{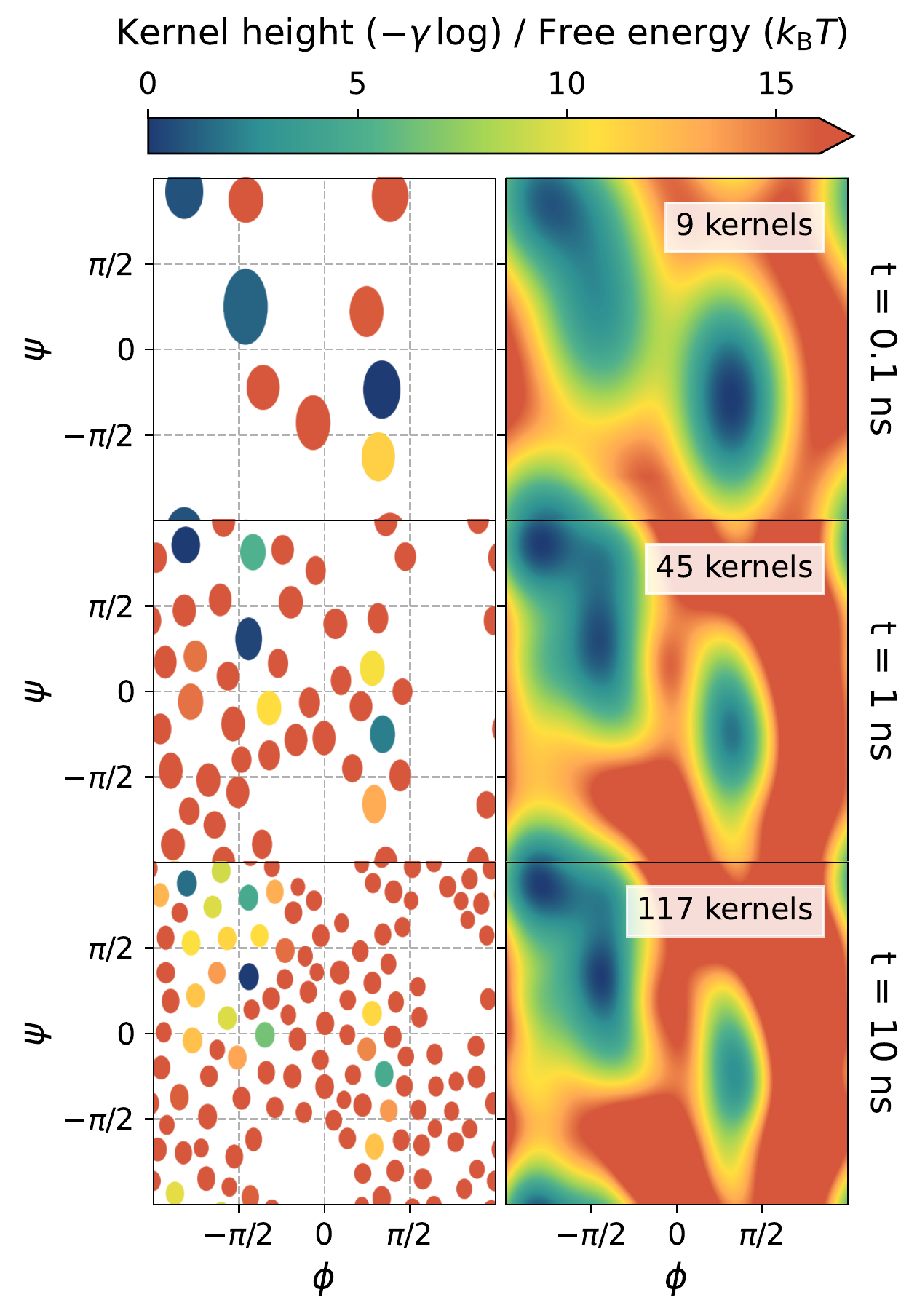} \\
        (a) OPES & (b) OPES EXPLORE
    \end{tabular}
    \caption{Time evolution of a typical simulation of alanine dipeptide in vacuum using the two OPES variants with the dihedral angles $\phi$ and $\psi$ as CVs.
    For each method, the compressed kernels are shown on the left with the point size indicating the adaptive bandwidth, and the corresponding free energy estimate $F_n(\phi,\psi)$ on the right.
    (a) In the original OPES, kernels make up the unbiased distribution estimate $P_n(\phi,\psi)$ and  $F_n(\phi,\psi)=-\frac{1}{\beta}\log P_n(\phi,\psi)$, while (b) in OPES-explore kernels make up the sampled distribution estimate $p^{\text{WT}}_n(\phi,\psi)$ and  $F_n(\phi,\psi)=-\gamma\frac{1}{\beta}\log p^{\text{WT}}_n(\phi,\psi)$.
    All $F_n(\phi,\psi)$ are shifted to have zero minimum.
    Notice how OPES-explore requires fewer kernels and visits higher FES regions.}
    \label{F:ala2-kernels}
\end{figure*}
In figure \ref{F:ala2-kernels} we contrast an OPES and OPES-explore simulation of alanine dipeptide in vacuum, which has become a standard test for enhanced sampling methods.
Both simulations have the same input parameters and use the adaptive bandwidth scheme described in the SI.
The bias is initially quite coarse, but the width of the kernels reduces as the simulation proceeds and the details of the FES are increasingly better described.
It can clearly be seen that the OPES-explore variant employs fewer kernels compared to the original OPES.
This is due to the fact that in OPES-explore the kernel density estimation is used for $p^{\text{WT}}(\mathbf{s}) \propto [P(\mathbf{s})]^{1/\gamma}$ that is a smoothed version of $P(\mathbf{s})$, and thus requires less details.
This more compact representation can be useful especially in higher dimensions, where the number of kernels can greatly increase despite the compression algorithm.
However, as a drawback it can result in a less accurate bias estimate, especially for large values of $\gamma$.

\section{Fewer transitions can lead to better convergence}
\begin{figure}
 \includegraphics[width=0.8\myfigwidth]{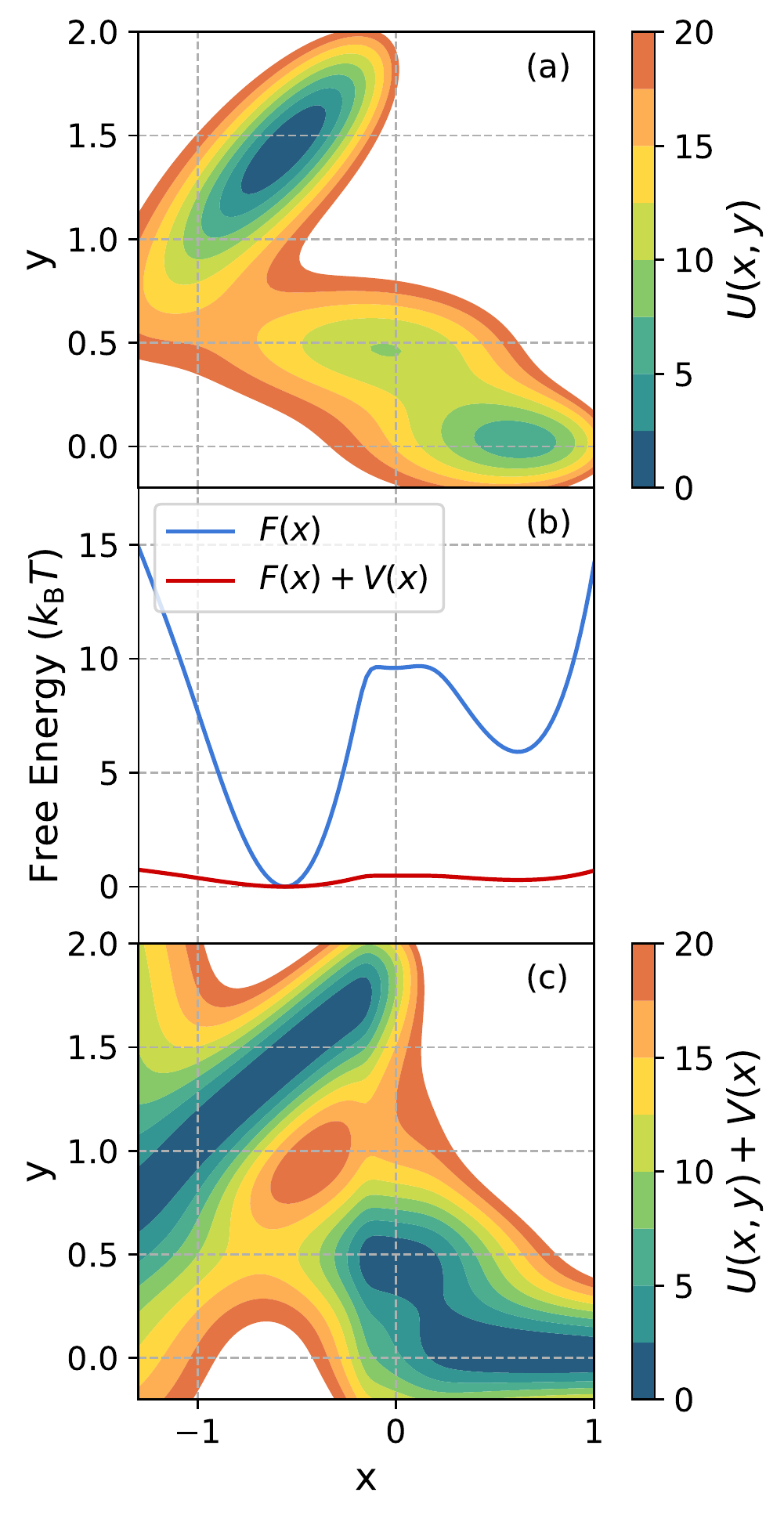}
  \caption{(a) The M\"uller potential energy surface, $U(x,y)$.
  (b) The free energy surface along the $x$ coordinate, $F(x)$, with and without the addition of the bias potential $V(x)=-(1-1/\gamma) F(x)$, where $\gamma=20$.
  (c) The potential energy modified by the bias potential, $U(x,y)+V(x)$.
  It can be seen that, despite the almost flat profile along $x$, the transition region between the states remains at high energy.
  }
  \label{F:mueller-potential}
\end{figure}
The difference in performance between OPES and OPES-explore cannot be judged from the alanine dipeptide example, because in this case the CVs chosen are extremely efficient.
In order to highlight the difference between the two methods, we study a simple two-dimensional model potential that is known as the M\"uller potential\cite{Muller1979}, see Fig.~\ref{F:mueller-potential}a, using the $x$ coordinate as collective variable.
This is a clear example of suboptimal CV, since it can discriminate the metastable states, but not the transition state.

For two-dimensional systems the free energy along the CV, $F(x)$, can be computed precisely with numerical integration, Fig.~\ref{F:mueller-potential}b.
From $F(x)$, the free energy difference between the two metastable states can be calculated as
\begin{equation}\label{E:mueller-deltaF}
    \Delta F = -\frac{1}{\beta} \log \frac{\int_0^1 e^{-\beta F(x)}dx}{\int_{-1.3}^0 e^{-\beta F(x)}dx}\, .
\end{equation}
While it is possible to distinguish better the two states by using also the $y$ coordinate, this does not result in a significant difference in the $\Delta F$ value (see SI, Sec.~S3).
On the other hand, $x$ does a poor job of identifying the transition state, which is around $x\approx -0.7$ and $y\approx 0.6$, and not at $x\approx 0$ as it would seem from $F(x)$.
As a consequence, it is not possible to significantly increase the transition rate between the states using a static bias that is a function of $x$ only.

To show this, we consider the effect of adding to the system the converged well-tempered bias $V(x)=-(1-1/\gamma)F(x)$, with $\gamma=20$.
In Fig.~\ref{F:mueller-potential}b, we can see the effect of the bias on the FES along $x$, which becomes almost completely flat.
However, when we consider the full 2D landscape, Fig.~\ref{F:mueller-potential}c, we can see that such bias does not really remove the barrier between the two states.
From the height of the barrier at the transition state, one can roughly estimate that adding $V(x)$ improves the transition rate of about one order of magnitude.
Nevertheless, transitions remain quite rare, around one every $10^6$ uncorrelated samples (see SI, Sec.~S3).

We want to compare the two OPES variants and well-tempered metadynamics in this challenging setting, where CVs are suboptimal and the total simulation time is not enough to reach full convergence.
This type of situation is not uncommon in practical applications, and it is thus of great interest.
Given enough time, all the method considered converge to the same bias potential and sample the same target distribution, but we shall see that  before reaching this limit they behave very differently.

\begin{figure*}
 \begin{tabular}{cc}
        \includegraphics[width=\myfigwidth]{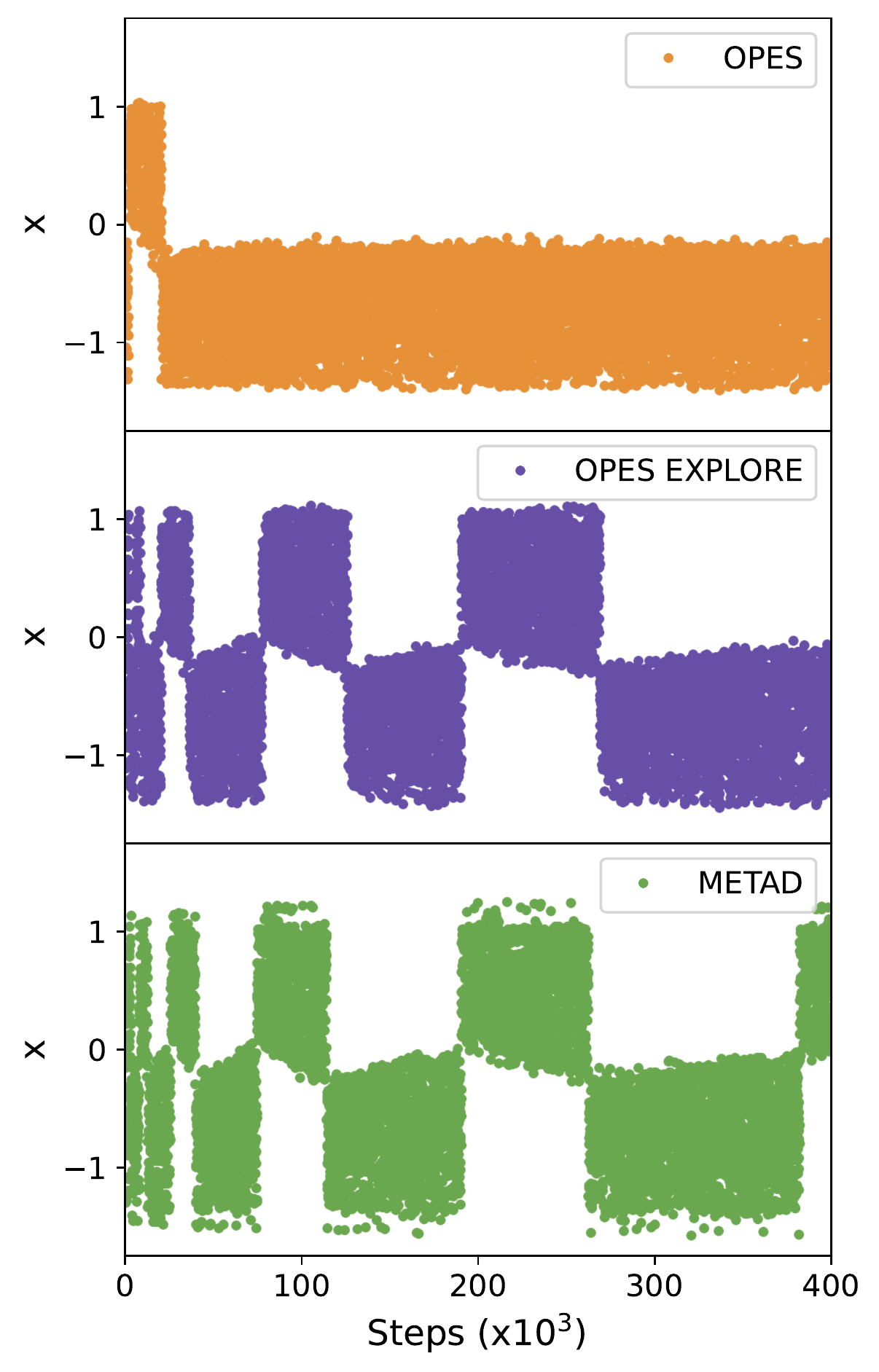}
        & \includegraphics[width=\myfigwidth]{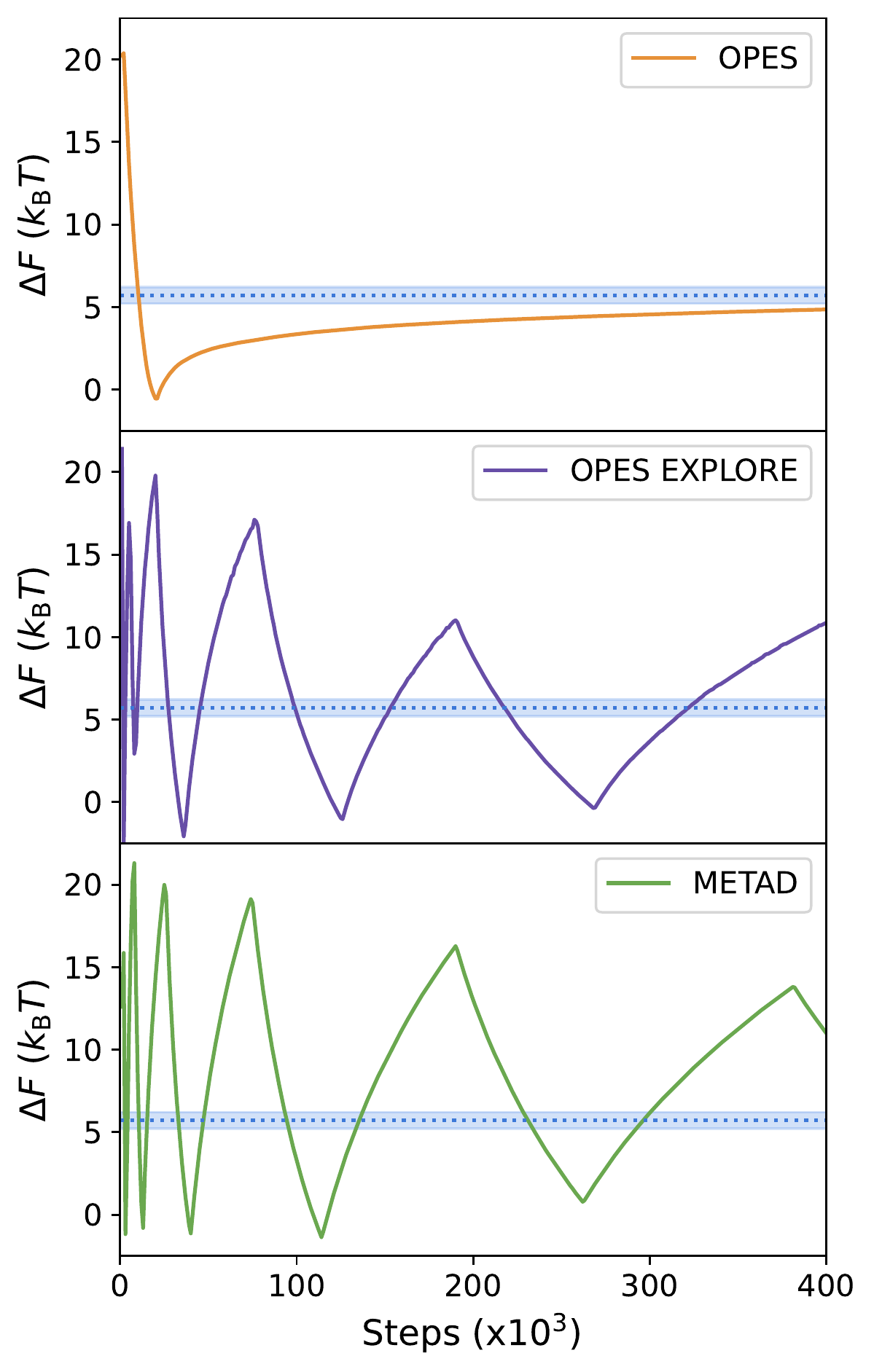} \\
        (a) $x$ trajectory & (b) $\Delta F$ estimate
    \end{tabular}
  \caption{Typical simulations of the M\"ueller potential using different methods for biasing the $x$ coordinate.
  Given more time, the three methods will converge to the same bias potential and will sample the same target distribution.
  In (a) is the trajectory along the CV and in (b) is the corresponding $\Delta F_n$, Eq.~\ref{E:mueller-deltaF}, calculated using the FES estimate obtained directly from the applied bias, $F_n(x)=-(1-1/\gamma)^{-1}V_n(x)$.
  The correct $\Delta F$ value is highlighted by a blue stripe 1~k$_\text{B}$T thick.
    }
  \label{F:mueller-single}
\end{figure*}
Figure \ref{F:mueller-single}a shows a typical run of the M\"uller potential obtained by biasing the $x$ coordinate with OPES, OPES-explore or MetaD.
As a simple way to visualize the evolution of the bias, we also report in Fig.~\ref{F:mueller-single}b the $\Delta F_n$ estimate obtained directly from the applied bias, by using $F_n(x)=-(1-1/\gamma)^{-1}V_n(x)$ in Eq.~(\ref{E:mueller-deltaF}).
We can see a qualitative difference between OPES and the other two methods.

OPES reaches a quasi-static bias that is very close to the converged one, but samples a distribution that is far from the well-tempered one, where the two basins would be about equally populated.
On the other hand, the $x$ distribution sampled by OPES-explore is closer to the target well-tempered one, but its bias is far from converged, and makes ample oscillations around the correct value.
Metadynamics behaves similarly to OPES-explore.
This is the exploration-convergence tradeoff described in the introduction.
Since the CV is suboptimal, even when using the converged bias $V(x)$, to see a transition occur one has to wait for an average number of steps $\tau \approx 10^6$, which is more than the total length of the simulation.
However, it is possible to greatly accelerate transitions by using a time-dependent bias that forces the system into higher energy pathways, that are not accessible at equilibrium.

In OPES-explore the bias is based on the estimate of the sampled probability $p^{\text{WT}}_n(\mathbf{s})$, and pushes to make it similar to the almost flat well-tempered target.
This means that in order to have a quasi-static bias, $p^{\text{WT}}_n(\mathbf{s})$ should both be almost flat and not change significantly as the simulation proceeds.
Clearly, this cannot happen unless the simulation is longer than $\tau$, otherwise most of the time would be spent in the same basin and $p^{\text{WT}}_n(\mathbf{s})$ would be far from flat.
On the contrary, in OPES the bias is based on the reweighted estimate $P_n(\mathbf{s})$, and thus it can reach a quasi-static regime even before sampling the target distribution.

\begin{figure*}
 \begin{tabular}{cc}
        \includegraphics[width=\myfigwidth]{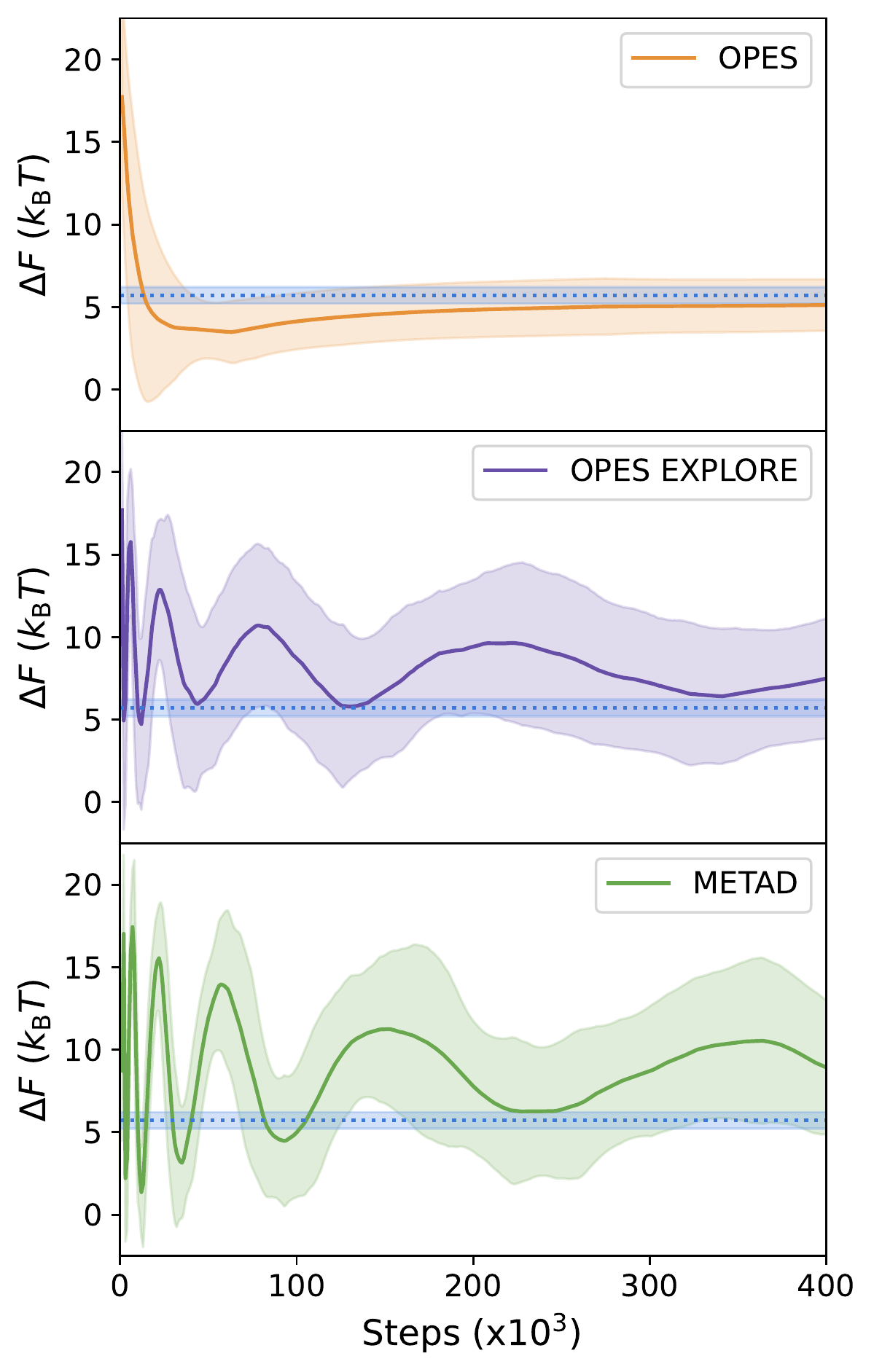}
        & \includegraphics[width=\myfigwidth]{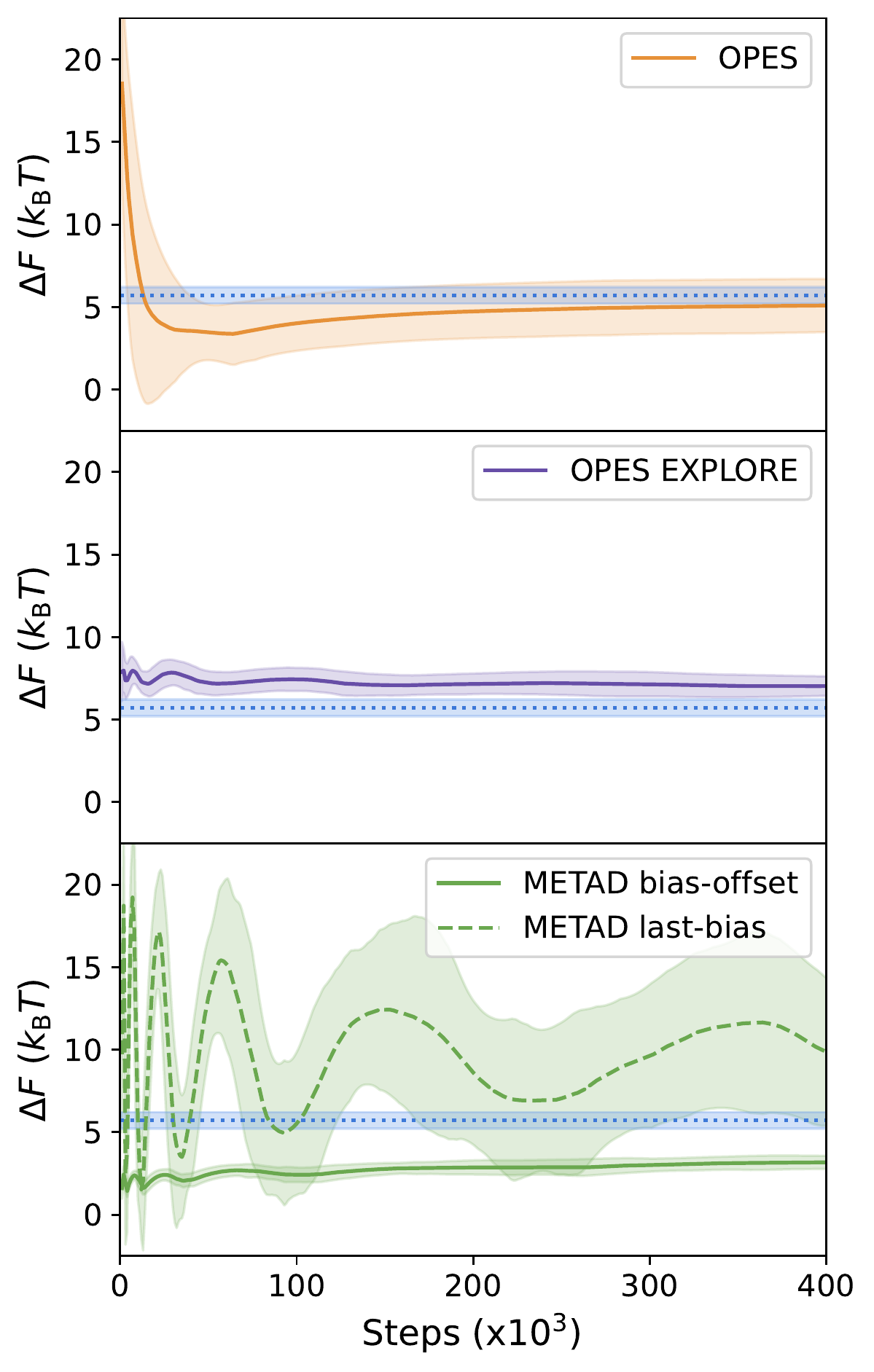} \\
        (a) Direct $\Delta F$ & (b) Reweighted $\Delta F$
    \end{tabular}
  \caption{Estimate of the free energy difference $\Delta F$ for the M\"ueller potential obtained by averaging 25 independent runs for each biasing method.
  The standard deviation is also shown for each estimate.
  Given more time, all these estimates will converge to the correct $\Delta F$.
  All simulations start from the main basin, $x<0$, but with different initial conditions.
  In (a) is the estimate obtained directly from the applied bias, as in Fig.~\ref{F:mueller-single}b, while in (b) is the corresponding estimate obtained via reweighting.
  For metadynamics two different reweighting schemes are considered, bias-offset \cite{Tiwary2015,Valsson2016} and last-bias reweighting\cite{Branduardi2012,Bussi2020}.
  The correct $\Delta F$ value is highlighted by a blue stripe 1~k$_\text{B}$T thick.
    }
  \label{F:mueller-average}
\end{figure*}
In figure \ref{F:mueller-average}a we show the $\Delta F_n$ estimate averaged over 25 independent runs, all starting from the main basin $x<0$.
We can see that on average OPES provides the best $\Delta F_n$ estimate at any $n$ in spite of the fact that it induces far less transitions.
In fact, most of the time only one full back-and-forth transition is observed (see SI).
One should notice that after a single transition the $\Delta F_n$ estimate is far from being accurate (see Fig.~\ref{F:mueller-single}b) but, since the bias quickly becomes quasi-static, it is possible to collect equilibrium samples and reliably reweight them, and the average estimate becomes more accurate the more simulations are run.
Instead in OPES-explore and MetaD, despite starting from independent initial conditions, the runs are highly correlated, due to the transitions being mostly driven by the strong changes in the bias rather than the natural fluctuations of the system.
As a further consequence of this, a systematic error is present in the average estimate, even if $\Delta F_n$ is further averaged over time, to remove the oscillatory behaviour of OPES-explore and MetaD.
Such systematic error depends on the characteristic of the system and the chosen CVs, and is hard to predict weather it will be relevant or small.
Nevertheless, one can be sure that it reduces over time as the bias converges\cite{Dama2014_convergence}.

Estimates of $\Delta F_n$ using different reweighting schemes are shown in Fig.~\ref{F:mueller-average}b.
For OPES and OPES-explore the simple Eq.~(\ref{E:reweighting}) has been used, while for MetaD we consider two of the most popular reweighiting schemes, namely last-bias reweighting\cite{Branduardi2012,Bussi2020} and bias-offset reweighting\cite{Tiwary2015,Valsson2016}.
As expected, the reweighting estimate of OPES is virtually identical to the direct estimate obtained from the bias, while for the other two methods the two estimates differ.
The reweighing of OPES-explore has very small statistical uncertainty, which further highlights the presence of a systematic error in the free energy difference estimate.
Like others before us\cite{Valsson2016,Salvalaglio2019,Giberti2020}, we observe empirically that the last-bias reweighting for MetaD tends to always be in agreement with the direct estimate, even when the simulation is far from converged, while the bias-offset reweighting provides a very unreliable estimate if the MetaD bias has not reached a quasi-static regime and the initial part of the simulation is not discarded.
Once again, it must be noted that the simulations considered here are not fully converged, otherwise all the different estimates of the various methods would have yielded the correct result, without systematic errors.
However, for most practical purposes they behave very differently, thus  it is important to choose between an exploration-focused or a convergence-focused enhanced sampling method, depending on the specific aim of the simulation.

\section{Sometimes exploration is what matters}
In the examples of the previous paragraph, it was shown in  that OPES
converges to a quasi-static bias faster than OPES-explore and provides more accurate FES estimates.
However, FES estimation is not the only goal of an enhanced sampling simulation.
In complex systems where good CVs are not available, convergence can remain out of reach, still one might be interested in exploring the phase space and find all the relevant metastable basins.
In such situation, OPES-explore can be a useful tool.

\begin{figure}
 \includegraphics[width=\myfigwidth]{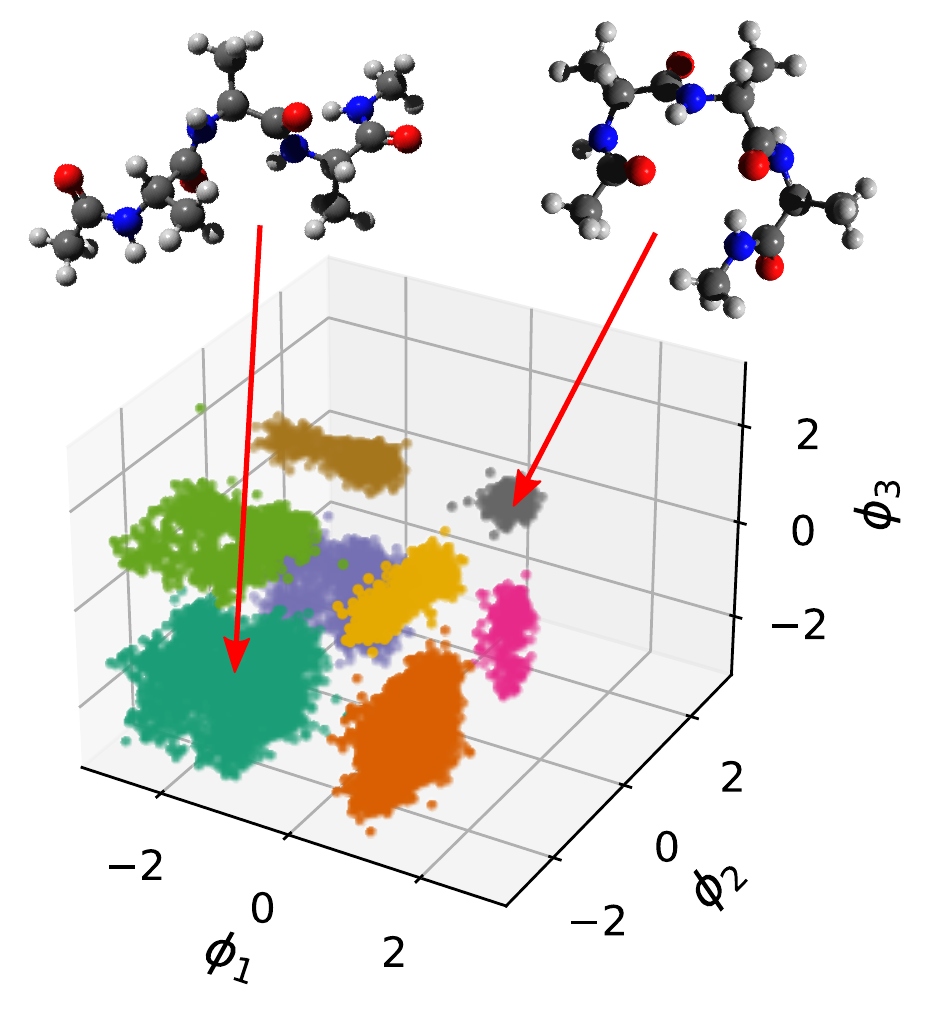}
  \caption{The eight metastable basins of alanine tetrapeptide in vacuum sampled via OPES-explore by biasing the three $\psi$ angles, a suboptimal set of CVs.
  Each basin is identified by the sign of the three $\phi$ angles, for a total of $2^3$ possible combinations.
  The most stable basin has $\phi_1,\phi_2,\phi_3<0$, while for the least stable $\phi_1,\phi_2,\phi_3>0$.
    }
  \label{F:ala4-phi_space}
\end{figure}
\begin{figure}
    \includegraphics[width=\myfigwidth]{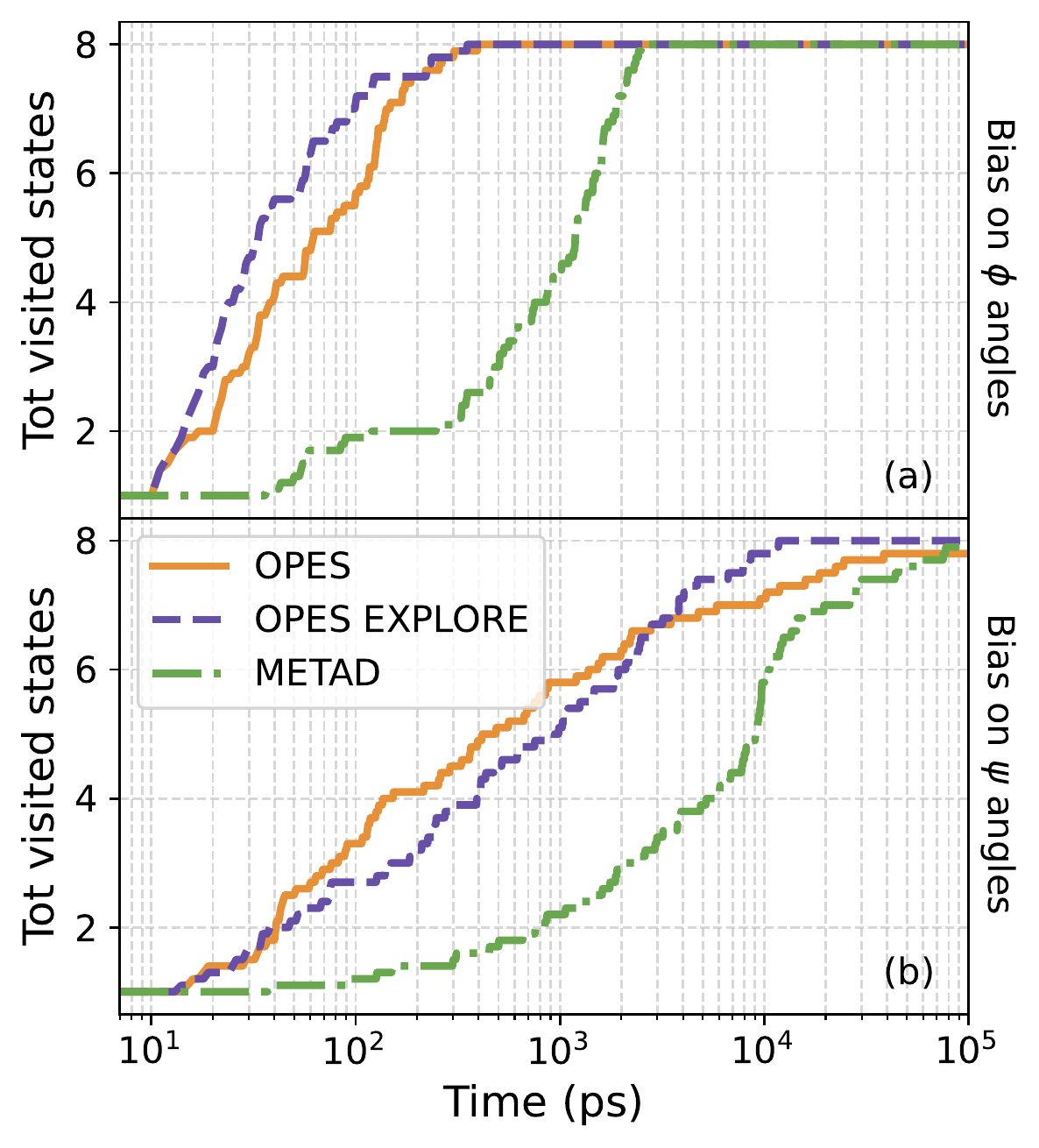}
    \caption{Exploration time of the eight metastable basins of alanine tetrapeptide over 100~ns.
    The lines are an average over 10 independent runs for each method, showing the total number of visited basins.
    In (a) the bias is a function of the three $\phi$ angles, $V=V(\phi_1,\phi_2,\phi_3)$, while in (b) the three $\psi$ angles are used, $V=V(\psi_1,\psi_2,\psi_3)$.
    See SI for results with different input parameters and other MetaD variants, such as parallel-bias MetaD\cite{Pfaendtner2015}.
    }
    \label{F:ala4-states}
\end{figure}
We consider here as test system alanine tetrapeptide in vacuum, as in Ref.~\citenum{Invernizzi2020rethinking}.
It has three $\phi$ dihedral angles, each of them can change  from positive to negative values and vice versa with a relatively low probability.
This leads to $2^3=8$ distinct metastable basins, each corresponding to a different combination of $\phi$ angles signs, as shown in Fig.~\ref{F:ala4-phi_space}.
Here we are not interested in estimating the FES, but rather we want to compare the ability of different methods to explore this space and discover all metastable states.

Figure \ref{F:ala4-states} shows the number of explored basins averaged over 10 independent simulations for each enhanced sampling method.
The simulations in the top panel (Fig.~\ref{F:ala4-states}a) use as CVs the $\phi$ angles, $V=V(\phi_1, \phi_2, \phi_3)$, which are good CVs, while in the bottom (Fig.~\ref{F:ala4-states}b) the suboptimal $\psi$ angles are used, $V=V(\psi_1, \psi_2, \psi_3)$.
In all methods, the exploration time increases approximately by two orders of magnitude when suboptimal CVs are used (please note the horizontal logarithmic scale).
As expected, OPES and OPES-explore have similar exploration speed when using good CVs, while with suboptimal CVs OPES struggles to find all the metastable basins.
This is because the same region of CV space might correspond to two different metastable basins, or to a basin and a transition state, as for the M\"uller potential\cite{Pietrucci2017,Bussi2020}.
In this situation, the previously estimated bias must change considerably for the simulation to escape quickly the current metastable state.

The exploration speed of MetaD depends critically on the input parameters and requires a trial-and-error tuning.
We report here only the outcome of MetaD simulations in which a standard choice of the input parameters has been made. 
As can be seen in Fig.~\ref{F:ala4-states}, in these simulations the exploration speed is roughly one order of magnitude slower than that of OPES-explore.
However, the performance of MetaD simulations can be improved by using different settings, as shown in the SI.
In the SI we also report and briefly discuss results obtained with non-tempered metadynamics\cite{Laio2002}, adaptive-Gaussians metadynamics\cite{Branduardi2012} and parallel-bias metadynamics\cite{Pfaendtner2015}.
None of these MetaD variants significantly improve the exploration speed, and some make it even worse.

Finally, in the SI we show how a preliminary OPES-explore run can be combined with a multithermal OPES simulation\cite{Invernizzi2020unified,Invernizzi2021} to sample efficiently alanine tetrapeptide and reach a converged FES, even without explicitly biasing the $\phi$ angles.

\section{Conclusion}
We have shown with the help of model systems that there is an exploration-convergence tradeoff in adaptive-bias methods when suboptimal CVs are used.
This tradeoff should not be confused with the exit time problem, that is present also with optimal CVs, and is discussed in Sec.~\ref{S:opes} and Refs.~\citenum{Dama2014_ttmetad,Fort2017}.
Contrary to the exit time problem, the exploration-convergence tradeoff cannot be solved and is an intrinsic limitation of CV-based adaptive-bias methods, that is a consequence of suboptimal CVs.
We believe the best way to handle this tradeoff is to have separate methods that clearly focus on one or the other aspect, so that they can be used depending on the application.
In a convergence-focused method the bias soon becomes quasi-static to allow for accurate reweighting and free energy estimation.
However, with suboptimal CVs this leads to a slow transition rate and a long time is required to sample the target distribution.
As discussed, even if one knows the true $F(\mathbf{s})$ and directly applies the converged bias, one would not obtain a faster exploration.
In an exploration-focused method, it is possible to improve the exploration speed by letting the bias change substantially even in a CV region that has already been visited.
While this may increase the number of transitions, it comes at the cost of a less accurate estimate of the free energy.

The original OPES method focuses on fast convergence to provide an accurate estimate of the free energy surface and reweighted observables.
As a consequence, it is very sensitive to the quality of the CVs (see e.g.~Fig.~\ref{F:mueller-single}a) and any improvement in the CVs results in a clear acceleration of the transition rate.
This is a particularly useful property when developing machine learning-based CVs, and in fact OPES has already been used several times in this context\cite{Bonati2020,Bonati2021,Karmakar2021,Trizio2021,Rizzi2021,Ansari2021}.

In other situations, improving the CVs may require first a better exploration of the phase space\cite{Branduardi2007,McCarty2017,Mendels2018,Bonati2021}.
Furthermore, one may be interested simply in exploring the metastable states of a system rather than estimating an accurate FES\cite{Piaggi2018,Capelli2019,Ahlawat2021,Francia2021}.
For this reason, we have introduced a variant of the OPES method, OPES-explore, that focuses on quickly sampling the target distribution and exploring the phase space.

We have shown that also well-tempered metadynamics is an exploration-focused method.
One of the main advantages of OPES-explore over MetaD is that it is easier to use, since it requires fewer input parameters and it has a more straightforward reweighting scheme (but more advanced ones can also be used\cite{Salvalaglio2019,Carli2021}).
Another important difference between the two methods is that OPES-explore, similarly to OPES, by default provides a maximum threshold to the applied bias potential, thus it avoids unreasonably high free energy regions.
To obtain the same effect with MetaD, one typically has to define some \textit{ad hoc} static bias walls by trial and error.
This last feature of OPES-explore has been recently leveraged by Raucci et al.~to systematically discover reaction pathways in chemical processes\cite{Raucci2022}.

Finally, we should clarify that OPES-explore, just as metadynamics, might not be able to exit any metastable state if the CVs are too poor\cite{Bussi2015,Bussi2020}, and its improved exploration capability can only be harnessed if the CVs are close enough to the correct ones to make such transitions possible.
The speed and small number of input parameters of OPES-explore are extremely helpful for quickly testing several candidate CVs, to find out which can drive transitions and discard the bad ones.

We believe that OPES-explore is an important addition to the OPES family of methods and will become a useful tool for researchers as it pushes forward the trend for more robust and reliable enhanced sampling methods.

\begin{acknowledgement}

We thank Valerio Rizzi and Umberto Raucci for useful discussions.
M.I. acknowledges support from the Swiss National Science Foundation through an Early Postdoc.Mobility fellowship.
Calculations were carried out on Euler cluster at ETH Zurich and on workstations provided by USI Lugano.

\end{acknowledgement}

\section*{Data availability}
An open-source implementation of the OPES and OPES-explore methods is available in the enhanced sampling library PLUMED from version 2.8\cite{plumed}.
All the data and input files needed to reproduce the simulations presented in this paper are available on PLUMED-NEST (\url{www.plumed-nest.org}), the public repository of the PLUMED consortium \cite{nest}, as plumID:22.003\,.

\begin{suppinfo}

Description of the adaptive bandwidth algorithm, computational details regarding the M\"uller potential and further biased trajectories, exploration speed for alanine tetrapeptide using other methods, and description of a multithermal-multiumbrella simulation to improve upon the OPES-explore run.

\end{suppinfo}

\bibliography{ref}

\newpage\hbox{}\thispagestyle{empty}\newpage 


\section*{Supplementary material}

# Supporting Information:
# Exploration vs Convergence Speed in Adaptive-bias Enhanced Sampling

## Michele Invernizzi[*,†] and Michele Parrinello[‡]

†*Freie Universität Berlin, 14195 Berlin, Germany*
‡*Italian Institute of Technology, 16163 Genova, Italy*

E-mail: michele.invernizzi@fu-berlin.de

## S1 Adaptive kernel bandwidth

Inspired by the adaptive Gaussian implementation of metadynamics,[S1] we introduce a similar feature that can be used in both OPES and OPES-explore. Typically, in OPES the initial kernel bandwidth $\boldsymbol{\sigma}^{(0)}$ (see Eq. (4) in the main text) is chosen as the square root of the CVs variance, obtained from a short unbiased run. We can however estimate its value adaptively during the simulation, by taking an exponentially decaying running average[S2]

$$\bar{\mathbf{s}}_m = \bar{\mathbf{s}}_{m-1} + (\mathbf{s}_m - \bar{\mathbf{s}}_{m-1})/\tau \qquad (S1)$$

$$\mathbf{S}_m = \mathbf{S}_{m-1} + (\mathbf{s}_m - \bar{\mathbf{s}}_{m-1})(\mathbf{s}_m - \bar{\mathbf{s}}_m) \qquad (S2)$$

Where the index $m$ indicates the simulation step, and should not be confused with the index $n$ that instead indicates the iteration step, or how many times the bias has been updated. Typically one iteration takes hundreds of simulation steps. The variance at simulation step $m$ is computed as $\mathbf{S}_m/m$. The default value for the parameter $\tau$ is set to 10 times one iteration pace.

Since what has been computed is the variance in the sampled ensemble, it can be directly used to set the initial kernel bandwidth in OPES-explore:

$$\boldsymbol{\sigma}^{(0)}{}_m = \sqrt{\mathbf{S}_m/m}\,, \qquad (S3)$$

In OPES instead, the kernels are used to reconstruct the unbiased probability density. Thus, the sampled variance must be rescaled by the bias factor $\gamma$:

$$\boldsymbol{\sigma}^{(0)}{}_m = \sqrt{\mathbf{S}_m/(\gamma m)}\,. \qquad (S4)$$

This expression is then used in Eq. (4) of the main text, with the caveat that for OPES-explore $N_{\text{eff}}^{(n)} = n$, since all kernels have same weight. Contrary to adaptive-Gaussians metadynamics,[S1] using the adaptive bandwidth in OPES does not introduce artifacts in the bias, and the self-consistency of the method is preserved.

The proposed adaptive bandwidth is still a global bandwidth, and does not locally adapt to the features of the underlying FES. For this reason, if a good guess of the initial kernel bandwidth $\boldsymbol{\sigma}^{(0)}$ is available, it can lead to similar or even better performance than when using an adaptive bandwidth. To this end, one can run a preliminary OPES-explore run with adaptive bandwidth and use it to estimate a good bandwidth value for a production run with OPES. The initial bandwidth should then be set to

$$\boldsymbol{\sigma}^{(0)} = \boldsymbol{\sigma}^{(0)}{}_m [n_m(d+2)/4]^{1/(d+4)}/\sqrt{\gamma}\,. \qquad (S5)$$

## S2 Simulation details

The OPES and OPES-explore methods are implemented in PLUMED v2.8[S3] and

can be found at `www.plumed.org/doc-v2.8/user-doc/html/_o_p_e_s.html`. The Langevin dynamics simulations with the Müller potential are carried out using PLUMED alone, while the alanine dipeptide and tetrapeptide simulations are done with GROMACS 2019[S4] compiled with the PLUMED plugin. All the input files for the simulations in the paper can be found on PLUMED-NEST[S5] at `www.plumed-nest.org/eggs/22/003` or alternatively at `www.github.com/invemichele/OPES-explore`.

## S3 Müller potential free energy difference

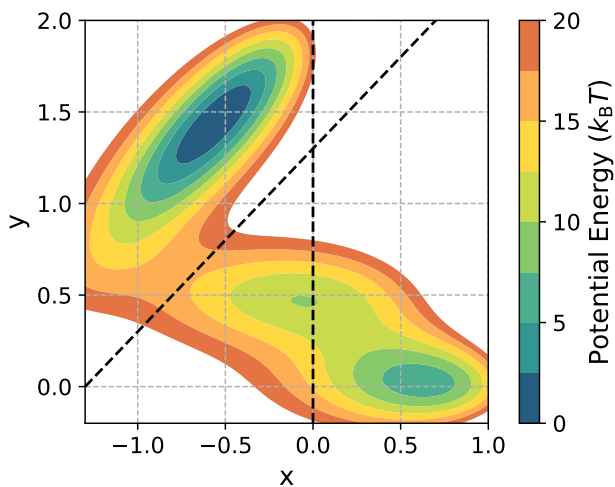

Figure S1: The Müller potential. The two dashed lines represent two possible partitions that define the metastable basins, one using only $x$ and the other (more accurate) using both $x$ and $y$. Both result in very similar $\Delta F$ values.

For simple 2D systems like the Müller potential (Fig. S1), it is possible to compute accurately the free energies, for instance via numerical integration. When calculating these free energies, both from numerical integration of from sampling biased simulations, we consider only the region of the plane shown in Fig. S1, thus $x \in [-1.3, 1.0]$ and $y \in [-0.2, 2.0]$. In the biased simulations we add a repulsive potential outside this region, to keep the simulation confined (see input files, Sec. S2). Thus, we con-

sider the following fee energy surface along $x$:

$$F(x) = -\frac{1}{\beta} \log \int_{-0.2}^{2.0} e^{-\beta U(x,y)} dy \,, \tag{S6}$$

while the free energy difference between the metastable basins A and B is:

$$\Delta F = -\frac{1}{\beta} \log \frac{\int_B e^{-\beta U(x,y)} dx \, dy}{\int_A e^{-\beta U(x,y)} dx \, dy} \,. \tag{S7}$$

In Eq. (10) of the main text, we have used the $x$ coordinate alone to identify the states, setting $A = \{(x,y) \mid x < 0\}$, and B by $B = \{(x,y) \mid x > 0\}$. With this definition one obtains that $\Delta F = 5.712(6)$. If the $y$ coordinate is also considered, one can define more accurately the two basins, for instance using as separator the line $y = x + 1.3$, see Fig. S1, thus $A = \{(x,y) \mid y > x + 1.3\}$ and $B = \{(x,y) \mid y < x + 1.3\}$. Even so, this results in a very similar $\Delta F = 5.690(5)$. Thus, although $x$ is a rather suboptimal CV, it can be used to estimate the free energy difference between the basins with good accuracy.

On the other hand, $x$ is quite bad when it comes to identify the transition state, and as a consequence it cannot be used to define a static bias that efficiently accelerates transitions between the metastable states. This is shown in Fig. 2c of the main text. When adding the converged well-tempered potential $V(x)$, the barrier at the transition state is only lowered from $\approx 16$ k$_B$T to $\approx 14$ k$_B$T. As a back of the envelope estimate, in the unbiased system we expect to sample the transition state once every $e^{16} \approx 10^7$ uncorrelated samples. Since to move from one state to the other one must pass through the transition state, this also gives a rough estimate of the transition rate. When running a simulation with the static bias $V(x)$, one should expect on average one transition every $e^{14} \approx 10^6$ uncorrelated samples, which is approximately what we find (see Fig. S5b). Note that "step" in x axis of these figures refers to the OPES (or MetaD) iterative step, that happens with a pace of 500 simulation steps, thus they can be approximately considered as loosely correlated samples.

Finally, for comparison we show the effect of

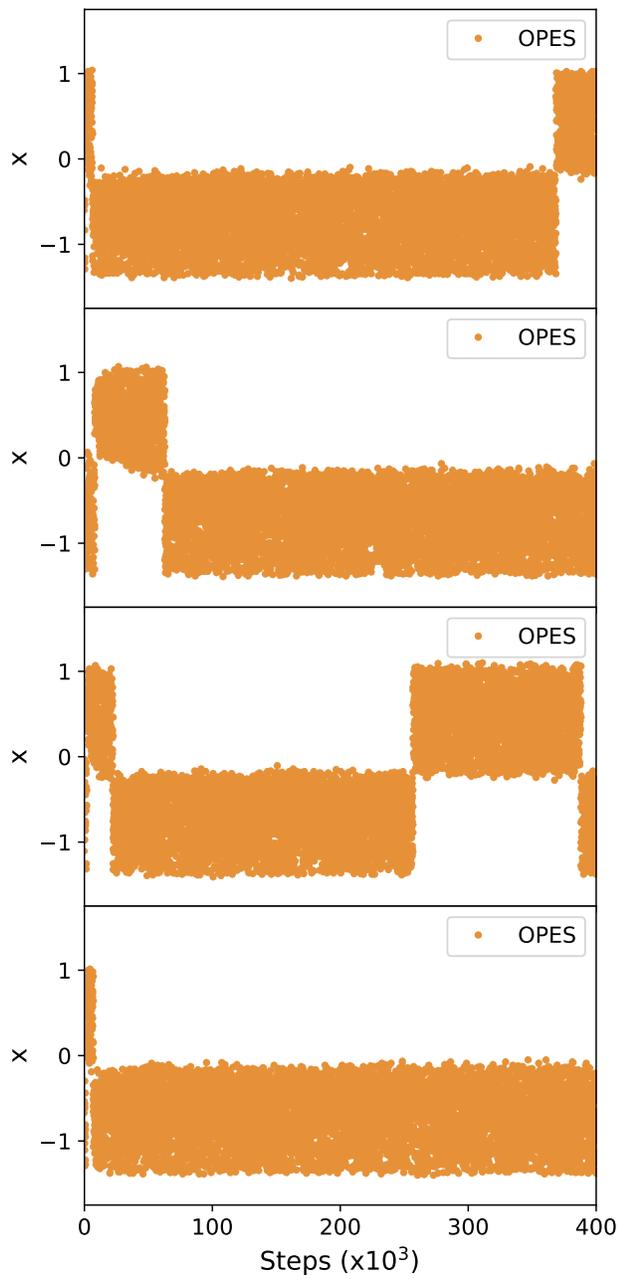

(a) $x$ trajectory

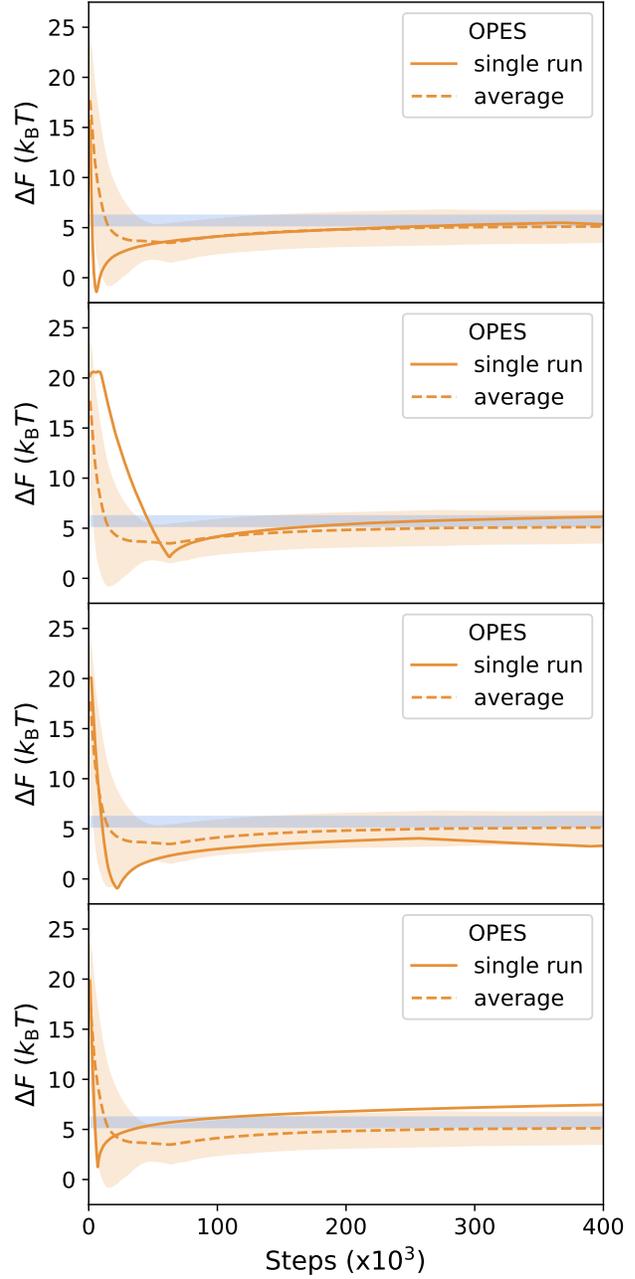

(b) $\Delta F$ estimate

Figure S2: Some other CV trajectories and corresponding estimates of free energy difference for the Müller potential, analogous of Fig. 3 of the main text. The average estimate over all the 25 runs is also shown for reference, same as Fig. 4a.

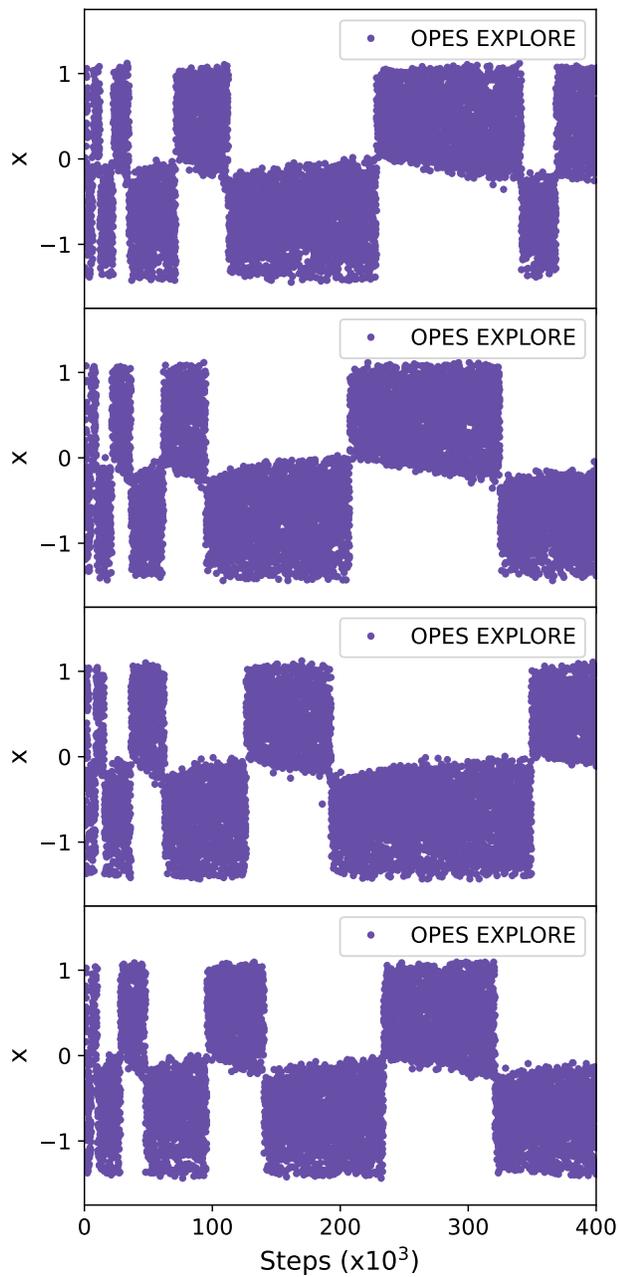

(a) $x$ trajectory

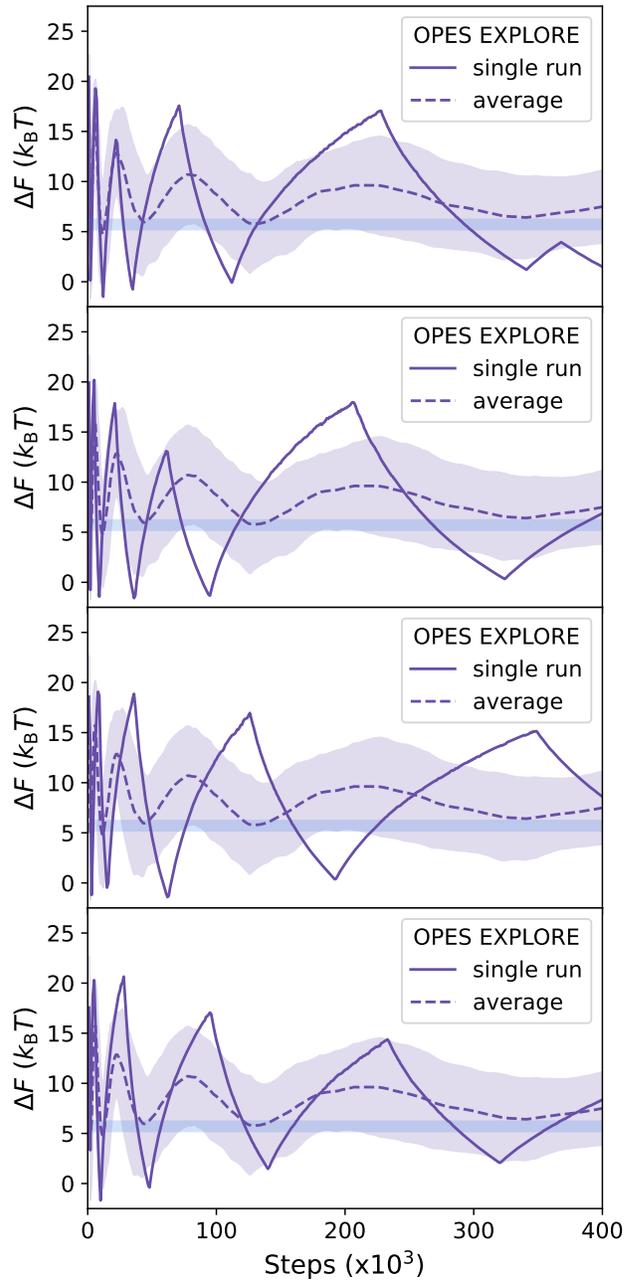

(b) $\Delta F$ estimate

Figure S3: Some other CV trajectories and corresponding estimates of free energy difference for the Müller potential, analogous of Fig. 3 of the main text. The average estimate over all the 25 runs is also shown for reference, same as Fig. 4a.

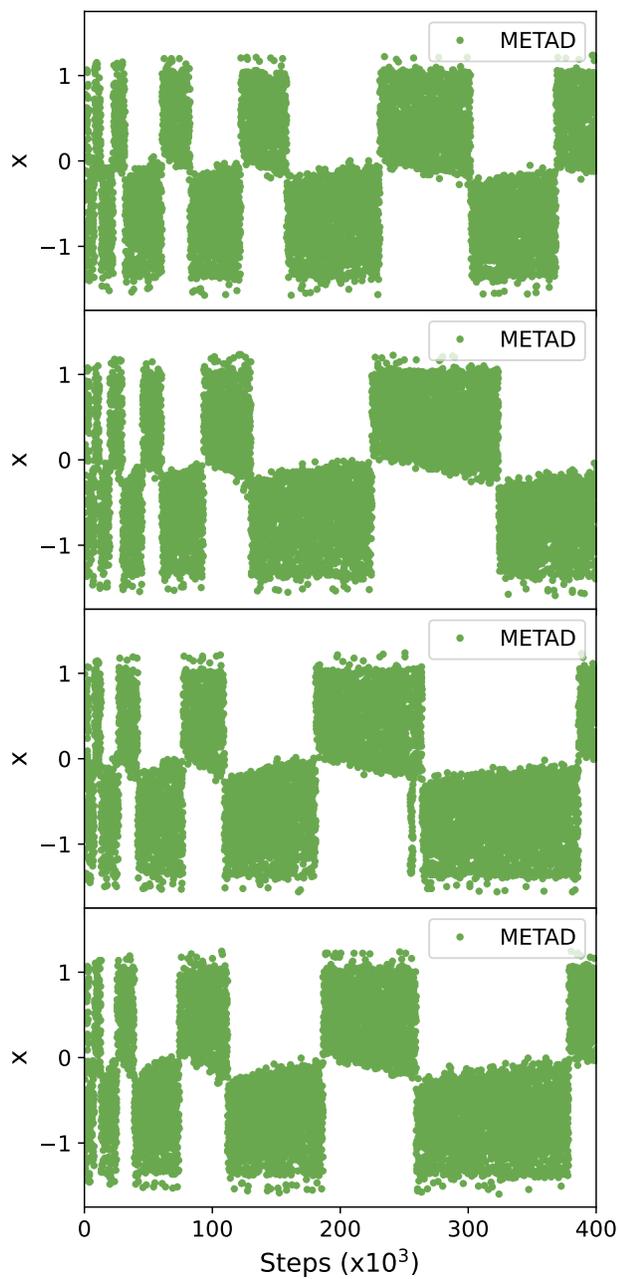

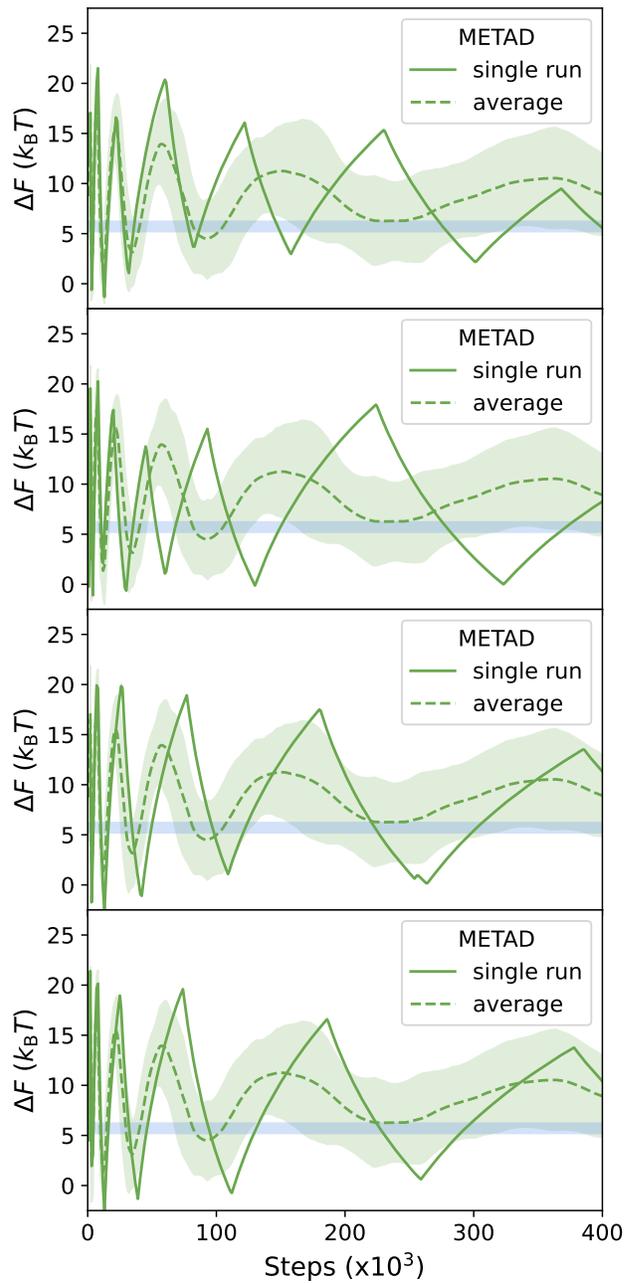

(a) $x$ trajectory

(b) $\Delta F$ estimate

Figure S4: Some other CV trajectories and corresponding estimates of free energy difference for the Müller potential, analogous of Fig. 3 of the main text. The average estimate over all the 25 runs is also shown for reference, same as Fig. 4a.

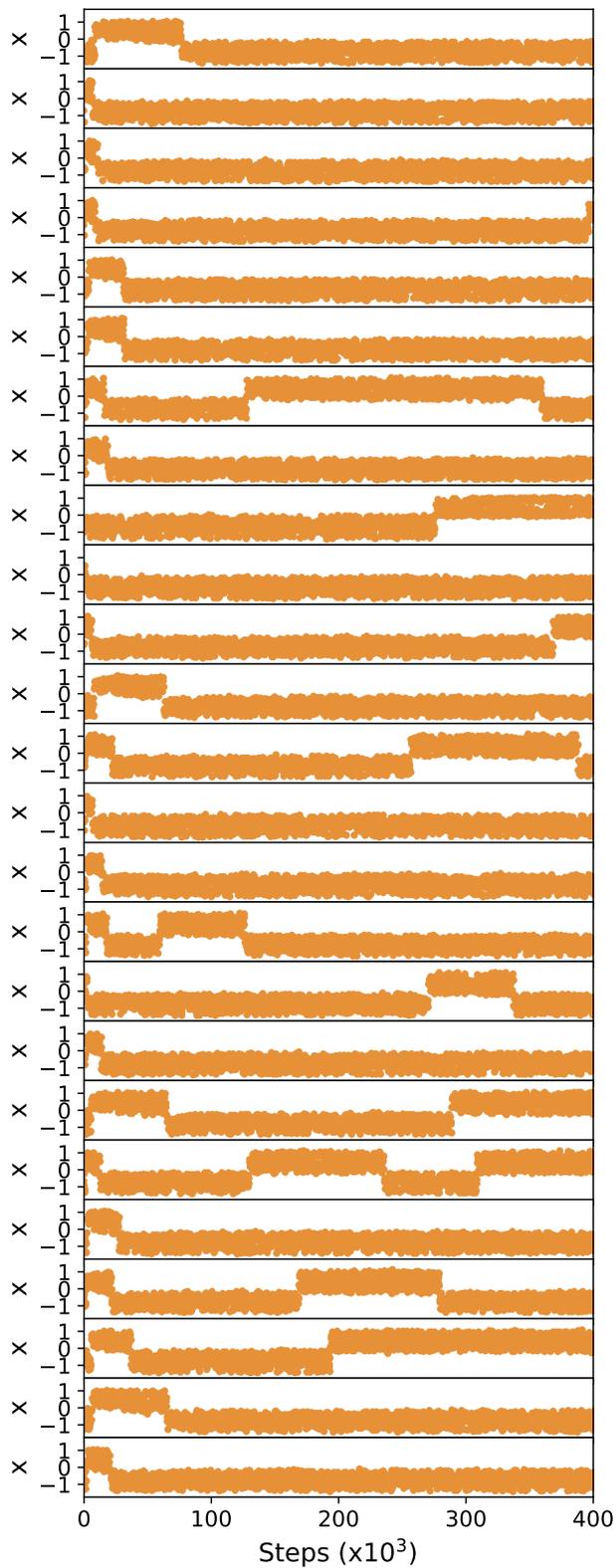

(a) OPES

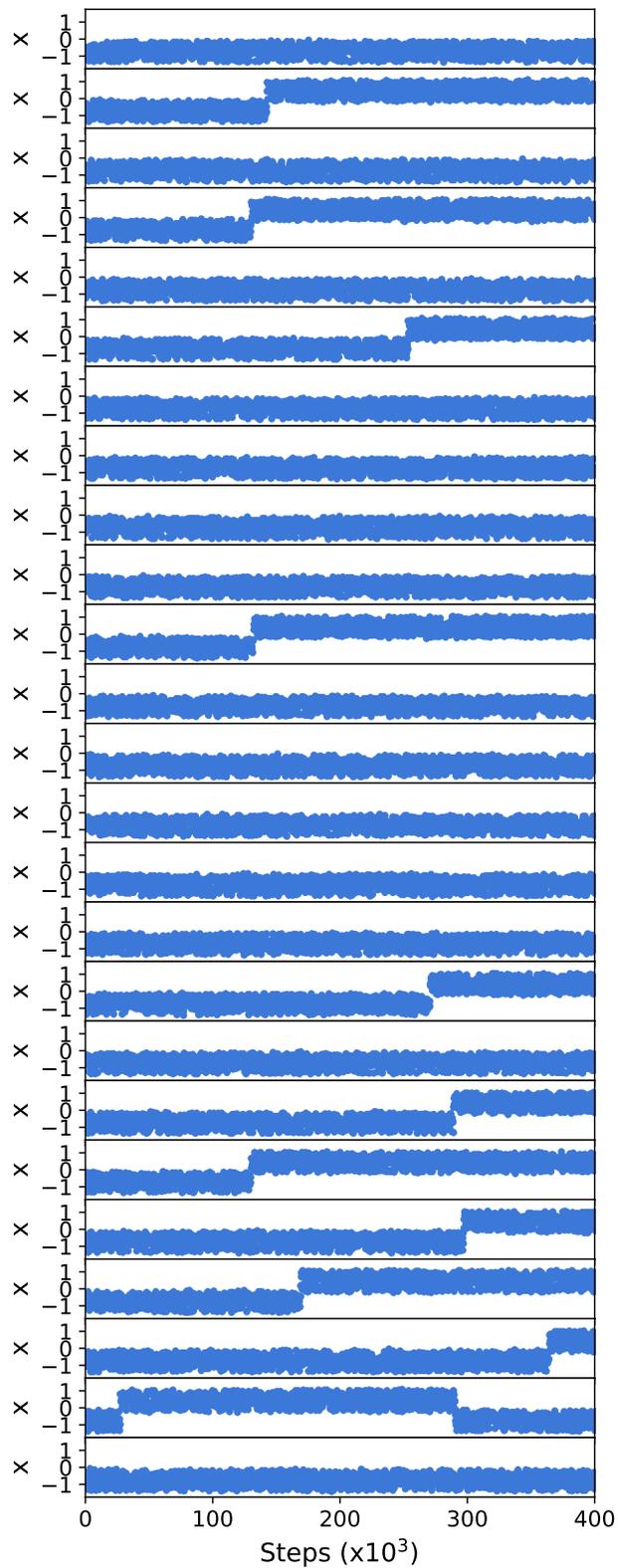

(b) Static bias

Figure S5: All the 25 independent trajectory of the Müller potential obtained by biasing (a) with the OPES method and (b) with the converged static bias $V(x) = -1(1 - 1/\gamma)F(x)$, thus sampling the potential energy shown in Fig. 2c of the main text.

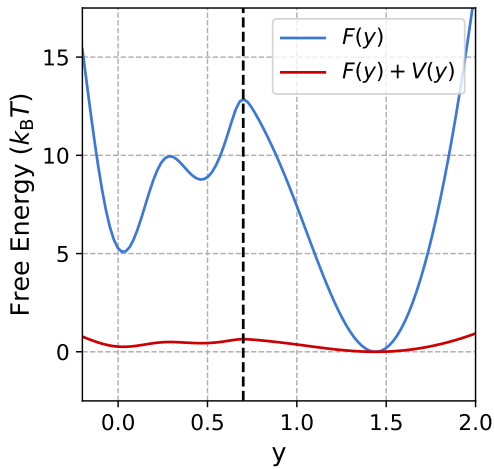

Figure S6: The free energy surface $F(y)$ of Müller potential along the $y$ coordinate. The effect of adding a converged well tempered potential $V(y) = -(1 - 1/\gamma)F(y)$ is also show, analogously to Fig. 2b of the main text.

biasing $y$ instead of $x$. The $y$ coordinate is a much better CV, and can identify the transition region as well as the metastable states. We can see from Fig. S6 that the barrier in $F(y)$ is higher than the one in $F(x)$, Fig. 2b of the main text, as expected when the CV describes a slower mode. In Fig. S7a we can see that when adding bias along $y$ the barrier is greatly reduced, contrary to what happens when biasing $x$, Fig. S7c or Fig. 2c of the main text.

## S4 Alanine tetrapeptide exploration

Alanine tetrapeptide in vacuum has eight distinct metastable states, identifiable by the values of its three $\phi$ dihedral angles. In figure 6 of the main text, we show the time required for visiting all these states using different enhanced sampling methods initialized with a standard choice of parameters. Here we further investigate the exploration time of other metadynamics variants, and different sets of input parameters. For all of the combinations, we tested the use of both good CVs (the three $\phi$ angles) and suboptimal CVs (the three $\psi$ angles). The deposition pace has not been investigated, and PACE = 500 is used in all the simulations. For

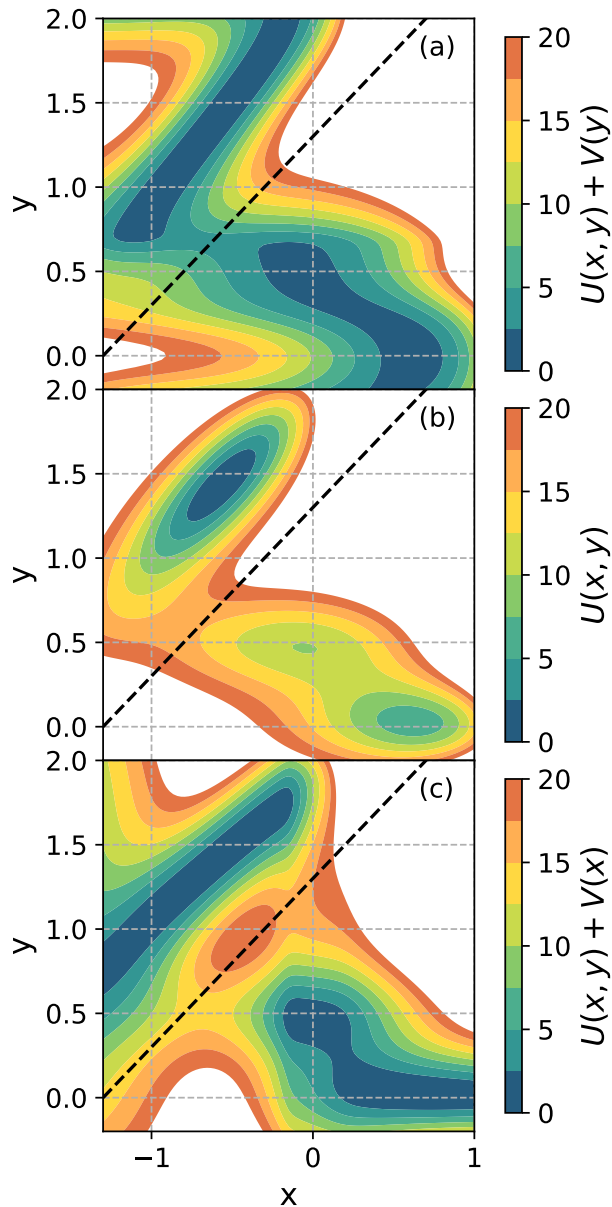

Figure S7: The effect of different bias potentials on the Müller system. In (a) the effect of biasing $y$ is shown, in (b) no bias is added (same as Fig. 2a of the main text) and in (c) the bias is added along $x$ (same as Fig. 2c of the main text). The barrier between the metastable states is much lower when a good CV like $y$ is biased, and this will result in much more transitions in a given simulation time.

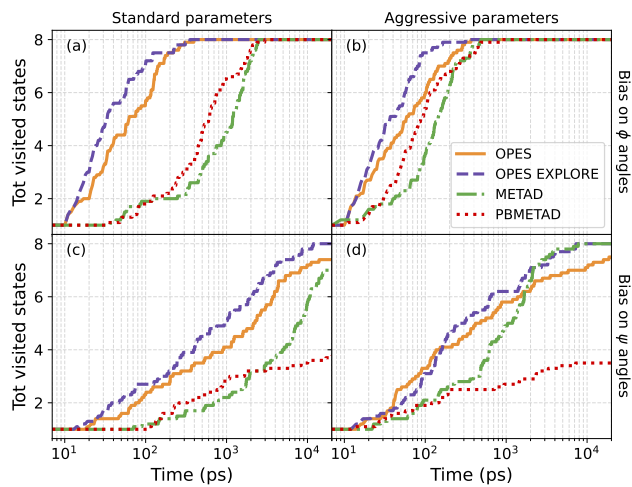

Figure S8: Exploration time of the eight metastable basins of alanine tetrapeptide over 20 ns. As for Fig. 6 of the main text, the lines are an average over 10 independent runs. In (a) and (b) the bias is a function of the three $\phi$ angles, while in (c) and (d) the three $\psi$ angles are used. See Sec. S4 for definition of "standard", (a)-(c), and "aggressive", (b)-(d), parameters.

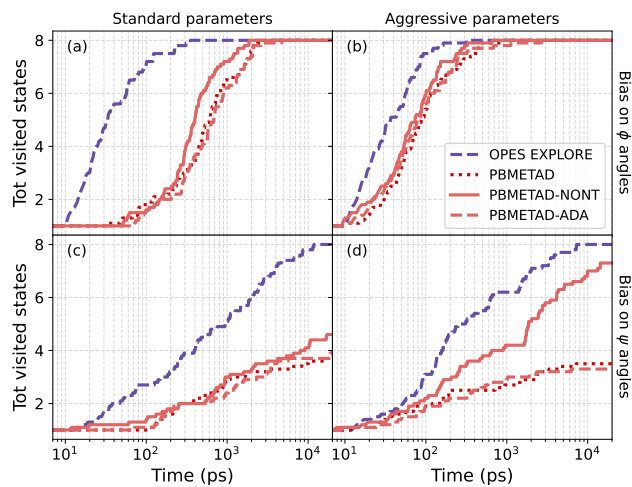

Figure S10: Same as Fig. S8, but with other methods. OPES-explore and PBMetaD are kept for reference, and non-tempered PBMetaD and adaptive-Gaussian PBMetaD are shown.

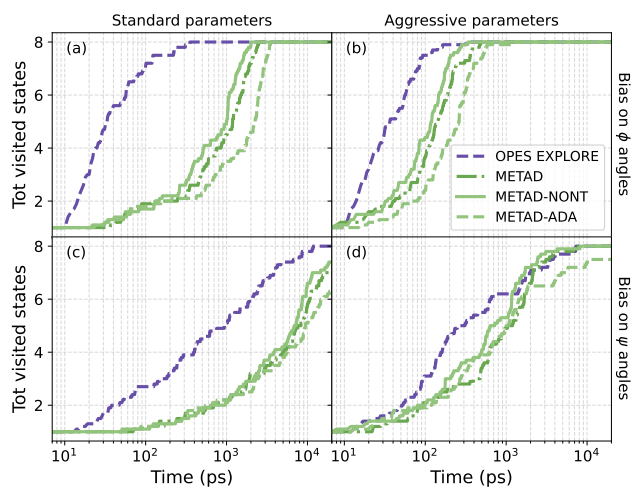

Figure S9: Same as Fig. S8, but with other methods. OPES-explore and MetaD are kept for reference, and non-tempered MetaD and adaptive-Gaussian MetaD are shown.

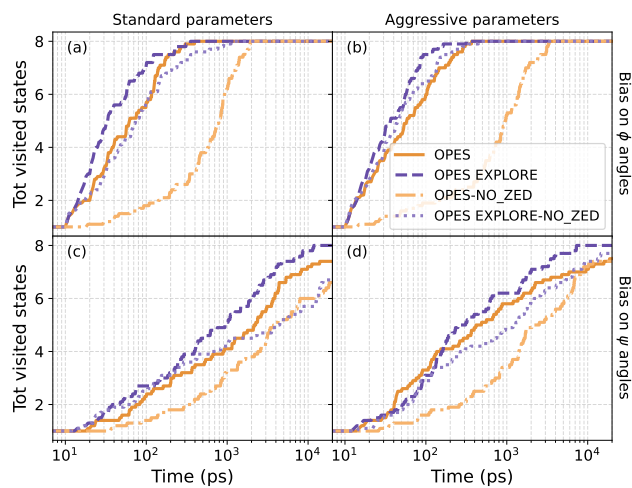

Figure S11: Same as Fig. S8, but with other methods. OPES and OPES-explore are kept for reference, and also a modified version without the $Z_n$ term is shown.

simplicity we label the parameters as "standard" or "aggressive". For OPES and OPES-explore the adaptive bandwidth algorithm is always used, and the only difference between "standard" and "aggressive" parameters is the value of the barrier, that is taken to be BARRIER = 50 and BARRIER = 70 respectively. We find that in metadynamics the most critical parameter for exploration speed is the initial Gaussian height, and choose HEIGHT = 1.2 as a "standard" value, and HEIGHT = 12 as "aggressive" value.

In figure S8, we show the same results of Fig. 6 of the main text, plus the popular parallel-bias metadynamics (PBMetaD).[S6] For both MetaD and PBMetaD we use BIASFACTOR = 20 and SIGMA = 0.35, 0.35, 0.35 . Figure S9 and S10 report different possible choices for these two parameters. We consider the non-tempered MetaD/PBMetaD[S7] (NONT) that corresponds to a bias factor $\gamma = \infty$ and it is known to not converge,[S8] and the adaptive-Gaussian MetaD/PBMetaD[S1] (ADA), with ADAPTIVE_SIGMA = DIFF and SIGMA = 500.

We can see from Fig. S9 that for MetaD the most critical parameter with respect to exploration speed is the Gaussian height. By increasing it of an order of magnitude compared to the usual value, we obtain an exploration speed similar to OPES-explore. However, it should be noted that a Gaussian height of 12 kJ/mol is a rather unusual choice for a biomolecule, and in other systems, e.g. with non-periodic CVs, such a high value would probably create instabilities and other difficulties. Perhaps surprisingly, increasing the bias factor up to non-tempered MetaD only brings a small improvement. This is due to the fact that the CV space is quite large (the bias is three-dimensional), and the simulation does not have time to return often to the regions already explored, so the well-tempered factor does not drastically reduce the hill heights. The use of an adaptive-Gaussian scheme always worsen the performance, but it is be the best available option in those cases where a good SIGMA value is not known.

Figure S10 reports the results for parallel-bias metadynamics, a variant of MetaD that is often used when dealing with high-dimensional CVs. The target distribution sampled by PBMetaD is the same of bias-exchange metadynamics,[S9] and it concentrates the sampling in a smaller volume of CV space when compared to the well-tempered distribution, allowing for higher efficiency. When biasing the $\phi$ variables, we find that PBMetaD has an exploration speed similar to MetaD, while its convergence speed improves (not shown). However, when biasing alanine tetrapeptide with the suboptimal $\psi$ angles, PBMetaD never manages to visit all the metastable basins, regardless of the chosen input parameters. We cannot say if this very slow exploration is due to the specific target distribution of PBMetaD or to the details of how the method works.

Finally, we report in Fig. S11 the effect of omitting the $Z_n$ term (see Eq. (5) and (8) of main text) in OPES and OPES-explore. It can be seen that the exploration speed worsens significantly, especially in the case of OPES with good CVs. This is a clear manifestation of the exit time problem discussed in the main text.

# S5    Alanine tetrapeptide convergence

While OPES-explore can be used to visit relatively quickly all the metastable basins even with the suboptimal $\psi$ CVs, it would require a much longer simulation to get a converged FES estimate. Nevertheless, it is possible to use the information obtained from the exploratory run, to perform another OPES simulation with a combined multithermal and $\psi$ bias, and quickly converge even without using the good $\phi$ CVs.

At higher temperatures, alanine tetrapeptide easily moves between those states that at 300K are metastable, and thus a multithermal simulation can be used to overcome the sampling problem. However, a multithermal simulation alone is not extremely efficient, and leaves room for improvement.[S10] Thus we want to combine a multithermal target distribution, with a multiumbrella distribution[S11] on the $\psi$ angles, where the centres and widths of the umbrellas are taken to be the same as the compressed

kernels obtained from the OPES-explore run. To sample this target distribution one could use an Hamiltonian replica exchange scheme,[S12] but it would require a very big number of parallel simulations. In the input files provided (Sec. S2) we show how this can be easily done with OPES[S11,S13] in a single simulation that can provide an efficient FES estimation even if the good $\phi$ CVs are unknown.

The advantage of using OPES with such expanded ensemble target instead of e.g. parallel tempering metadynamics[S14] or integrated tempering metadynamics,[S15] is a faster convergence of the reweighted observables, since those MetaD-based methods focus instead on quickly sampling the target distribution, and do not have straightforward reweighting schemes.



\end{document}